\documentclass[aps,prd,nofootinbib,preprint,superscriptaddress]{revtex4-2}
\usepackage{silence}
\usepackage{amsmath, amssymb, mathtools, bm, dcolumn, array, multirow}

\usepackage{graphicx}
\usepackage{xcolor}
\usepackage[compat=1.1.0]{tikz-feynman}
\tikzfeynmanset{warn luatex=false}

\usepackage{tabularray}
\UseTblrLibrary{booktabs}

\usepackage[font=small,labelfont=bf,compatibility=false]{caption}
\usepackage{subcaption}  

\allowdisplaybreaks
\usepackage[colorlinks=true,linkcolor=blue,citecolor=blue]{hyperref}

\usepackage{cleveref}

\usepackage{silence}
\begin{document}
\title{
Dark Matter–Driven Low-Scale Leptogenesis via Neutrino Portal
}
\author{Suresh Chand}
\email{sures176121102@iitg.ac.in}    
\affiliation{Department of Physics, Indian Institute of Technology Guwahati, Assam 781039, India}
\author{Avnish}
\email{avnish.yd@rnd.iitg.ac.in}
\affiliation{Department of Physics, Indian Institute of Technology Guwahati, Assam 781039, India}
\affiliation{Department of Physics, Indian Institute of Technology Guwahati, Assam 781039, India}
\author{Poulose Poulose}
\email{poulose@sju.edu.in}
\affiliation{Department of Physics, St. Joseph's University, Bengaluru 560 027, India}
\begin{abstract}
We propose a novel framework for low-scale leptogenesis within an extension of the Standard Model (SM) that includes three $SU(2)$ singlet right-handed neutrinos, a singlet charged neutral fermion, and a real scalar field. In this setup, the $C\!P$ asymmetry arises through a rich interplay of mechanisms, including two-body decays of the lightest right-handed neutrino into leptons and Higgs or into dark-sector particles, as well as multiple $2 \to 2$ scattering processes involving visible and dark states. Crucially, the $  C\!P$-violating phases originate not only from conventional vertex and self-energy corrections but also from novel interference effects mediated by the dark sector, which significantly enrich the sources of asymmetry. A distinctive feature of our model is the direct connection between the dark sector and leptogenesis, providing a unified explanation for both the matter–antimatter asymmetry and DM abundance. This connection leads to enhanced $C\!P$ violation in neutrino interactions and predicts new dark-sector particles accessible at the LHC. 
\end{abstract}

\maketitle
\newpage

\section{Introduction}
\label{sec:level1}
The Baryon Asymmetry of the Universe (BAU) is one of the most puzzling problems of modern cosmology. While the SM satisfies the qualitative requirements set through Sakharov Conditions~\cite{Sakharov:1967dj}, the experimentally measured Higgs mass renders the electroweak phase transition a smooth crossover~\cite {Dine:1991ck, PhysRevDDine, JAIN1993315, ESPINOSA1993206}, and the $C\!P$ violation present in the SM is far too small to  account for the amount of the BAU adequately. There are many beyond the Standard Model (BSM) frameworks have been proposed that invoke a first-order phase transition associated with electroweak symmetry breaking~\cite{Kuzmin:1985mm, Rubakov:1996vz, Dolgov:1991fr} to fulfill the required out-of-equilibrium condition. However, within the SM gauge symmetry, an alternative approach is considered. Rather than directly producing a baryon asymmetry, this approach first generates an asymmetry in the leptonic sector through lepton number ($L$)-violating decays. Subsequently, this lepton asymmetry can be converted into the observed baryon asymmetry through $(B+L)$-violating electroweak (EW) sphaleron processes~\cite{Kuzmin:1985mm}. This widely studied approach, known as \textit{leptogenesis}~\cite{Fukugita:1986hr}, was proposed by Fukugita and Yanagida. A comprehensive review of leptogenesis can be availed at Ref.~\cite{Davidson:2008bu}.
An appealing aspect of the standard leptogenesis is that the~  $\!C\!P$-violating, out-of-equilibrium decays responsible for generating lepton asymmetry can originate from the same heavy neutrinos involved in the seesaw mechanism~\cite{Minkowski:1977sc, Turner:1993dm,  Bertone:2016nfn, Falkowski:2017pss}, which explains the smallness of neutrino masses~\cite{ParticleDataGroup:2018ovx}. 

While matter-antimatter asymmetry remains one of the long-standing mysteries of the Universe, another intriguing puzzle lies in the composition of matter itself in the Universe. Observations from the spectral analysis of the cosmic microwave background (CMB) radiation \cite{Planck:2018vyg}, and many other independent observations 
\cite{Turner:1993dm, ArkaniHamed:2008qn, Bertone:2016nfn, Falkowski:2017pss} have shown that about $85\%$ of the matter content of the Universe is non-baryonic and non-luminous dark matter (DM). While dark matter plays a central role in cosmic structure formation through its gravitational interactions, it is usually expected to couple only very weakly to the Standard Model particles, as stronger interactions would have resulted in observable signals. Although neutrinos share several characteristics required for DM particles, their tiny mass and a large free-streaming length make them insufficient to fully account for the dark matter content of the Universe~\cite{Peters:2023asu}. Since none of other SM particles possess the required properties of a viable DM candidate, a plethora of beyond-the-Standard-Model (BSM) frameworks have been proposed to account for its existence.
Among the various DM paradigms, the weakly interacting massive particle (WIMP) scenario is one of the most extensively studied frameworks~\cite{Arcadi:2021doo}. In this picture, a DM particle with a mass around the electroweak scale and weak-scale interactions is thermally produced in the early Universe. As the Universe expands and cools, the interaction rate drops below the Hubble expansion rate, causing the dark matter to decouple from the thermal bath. The resulting frozen-out abundance naturally accounts for the observed dark matter relic density. 
However, conventional WIMP scenarios are being increasingly challenged by the difficulty of simultaneously satisfying the observed dark matter relic density and the stringent bounds from direct detection experiments ~\cite{Billard:2021uyg}. In direct detection searches, WIMPs are expected to elastically scatter off atomic nuclei, producing measurable nuclear recoil signals. In standard WIMP frameworks, the same interactions that determine the thermal relic abundance also control the DM–nucleus scattering cross section. Consequently, the null results from direct detection experiments impose severe constraints on the parameter space consistent with the observed relic density, significantly limiting the viability of such models. Motivated by these tensions, alternative paradigms have been proposed. 
One attractive scenarios among these is the so-called feebly interacting massive particles (FIMPs). In FIMP scenario, the dark matter interacts so weakly with the Standard Model that it is produced non-thermally in the early Universe and naturally evades direct detection bounds.
 This production mechanism, known as \textit{freeze-in}, leads to a relic density that asymptotically approaches the observed DM relic density~\cite{McDonald:2001vt, Hall:2009bx, Chand:2022vrf}. 
 Extensive analyses of the freeze-in mechanism have been explored in Ref.~\cite{Belanger:2018sti, Klasen:2013ypa, Elahi:2014fsa, Co:2015pka}, and specifically within the context of baryogenesis to explain the BAU in Ref.~\cite{Goudelis:2021qla}. 

In this work, we propose a unified framework that simultaneously addresses the origin of the baryon asymmetry of the Universe, the nature of dark matter, and the generation of neutrino masses. The model constitutes a minimal extension of the SM particle content, introducing a singlet vector-like fermion $\psi$, a neutral real scalar singlet $\phi$, and three generations of heavy right-handed neutrinos $N_{1,2,3}$. The dark sector is stabilized by a discrete $Z_2$ symmetry, under which $\psi$ and $\phi$ are odd, while the right-handed neutrinos and all the SM fields are even. In conventional type-I seesaw scenarios with hierarchical right-handed neutrinos, successful thermal leptogenesis requires the lightest heavy neutrino mass to satisfy the Davidson–Ibarra bound, $m_{N_1}\ge 10^9$ GeV \cite{Davidson:2002qv}. Various models  allowing low-scale leptogenesis with hierarchical neutrinos have been explored in the literature \cite{Hambye_2009, Racker_2014, PhysRevD.92.033006, PhysRevD.98.023020, PhysRevD.99.055012, Mahanta_2019, Mahanta_2020, SARMA2021115300}, while scenarios involving quasi-degenerate right-handed neutrinos, including resonant leptogenesis with enhanced $C\!P$ asymmetry are studied in the Refs. \cite{Pilaftsis:2003gt, Pilaftsis:2005rv, Pilaftsis:2009pk, PhysRevD.86.053001, europian, PhysRevD.86.053001, europian}. In contrast to these approaches, in the present setup, we demonstrate that a sufficient leptogenesis can be achieved with the lightest heavy neutrino mass being at the TeV scale without invoking resonant enhancement. This becomes possible through the neutrino portal interactions linking the right-handed neutrinos to the dark sector fields. These interactions open additional decay channels, loop contributions and scattering processes, leading to sizable dark-sector–induced 
$C\!P$ violation. Moreover, the new processes involving leptons and anti-leptons in both initial and final states modify the reaction rates entering the Boltzmann equations, thereby significantly impacting the evolution of the lepton asymmetry. A key feature of this scenario is the intimate interplay between dark matter dynamics and leptogenesis: the same couplings that govern the dark sector phenomenology also play a central role in enhancing the $C\!P$ asymmetry required for successful BAU generation. We show that the region of parameter space consistent with the observed dark matter relic abundance and other phenomenological constraints naturally accommodates the required low-scale leptogenesis. Furthermore, the presence of $C\!P$-violating interactions in the dark sector can also lead to an asymmetry within the dark sector itself, opening up additional avenues for phenomenological exploration.

The structure of the article is as follows. In Section ~\ref{sec:model}, we present the theoretical framework of the model, including the particle content, symmetries, and relevant interactions. Section~\ref{sec:dm} is devoted to the dark matter phenomenology, where the WIMP  realizations is discussed and the parameter space consistent with the observed relic abundance is identified. In Section~\ref{sec:leptogenesis}, we analyze the mechanism of leptogenesis, highlighting the novel sources of $C\!P$ violation arising from neutrino–portal interactions with the dark sector, including decay and scattering contributions. The numerical solutions of the coupled Boltzmann equations and a detailed discussion of the generated asymmetries are presented in Section~\ref{sec:numericalanalysis}. Finally, we summarize our results and outline future directions in Section~\ref{sec:relultandconclusion}.
\section{\label{sec:model} The Model}
The model framework under consideration extends the Standard Model (SM) by introducing three right-handed gauge-singlet neutrinos, denoted by $N_i$, a singlet vector-like fermion $\psi$, and a singlet real scalar field $\phi$. 
Both $\phi$ and $\psi$ are assigned to be odd under a discrete $Z_2$ symmetry, while all other fields remain even. This discrete symmetry is introduced to ensure the stability of the dark matter (DM) candidate. 
The gauge symmetry of the model is given by the SM gauge group supplemented with an additional discrete $Z_2$ symmetry. Depending on the mass hierarchy between $\phi$ and $\psi$, the lighter of the two becomes stable and can serve as a viable DM candidate. In addition, we assign a generalized lepton number (global $U(1)_{L'}$ symmetry) to the fermion $\psi$, the heavy neutrinos $N_i$, the SM lepton doublet $L$ and the SM lepton singlet $e_R$, while the scalar $\phi$ and rest of the Standard Model particles are taken to be lepton-number neutral. This symmetry governs the generation of asymmetries by distinguishing $\psi$ from its antiparticle $\bar{\psi}$. The relevant particle content of this extended model, along with their transformation properties under the SM gauge symmetry, $U(1)_{L'}$ symmetry, and the $Z_2$ symmetry, is summarized in Table~\ref{tab:field1}. 
\begin{table}[!htb]
\begin{center}
\begin{tabular}{ c | c|c | c | c | c |c}  
     \hline \hline
      ~Particles~ & Spin&~${\rm SU(3)}_c$~ &~${\rm SU}(2)_{L}$~  &~${\rm  U}(1)_{Y}$~  &$U(1)_{L'}$& ~${\rm Z}_{2}$ ~ \\
      \hline 
      $N_{1,2,3}$ & $\tfrac{1}{2}$& 1 &1  &0  & 1&1     \\
      $\psi$& $\tfrac{1}{2}$      &1  &1  &0  &1& -1   \\
       $\phi$  & 0   &1  &1  &0               &0& -1      \\
      $L$  & $\tfrac{1}{2}$ &1  &2  & $-\tfrac{1}{2}$   &1 &1      \\
      $e_R$  & $\tfrac{1}{2}$   &1  &1  &-1   &1 &1      \\
       \hline \hline
\end{tabular}
\end{center}
\caption{The relevant particle content of the model. All the SM particles are $Z_2$ even. The SM leptons have $U(1)_{L'}$ charge equal to $+1$, while all the other SM particles  have zero $U(1)_{L'}$ charge.}
\label{tab:field1}
\end{table} 

The Lagrangian for this model framework is given by
\begin{align}
\mathcal{L}_m \;=\;& \mathcal{L}_{\rm SM}
+ (\partial^\mu \phi)^\dagger (\partial_\mu \phi)
+ \bar{\psi}\, i \gamma^\mu \partial_\mu \psi
+ \bar{N}_i\, i \gamma^\mu \partial_\mu N_i
- m_\psi\, \bar{\psi}\psi
- m_{N_{ij}}\, \bar{N}_i N_j
\nonumber\\[2pt]
&- \left(
y_i\, \bar{\psi}\,\phi\, N_i
+ Y^N_{ij}\, \bar{L}_i \tilde{H} N_j
+ {\rm h.c.}
\right)
- \lambda'\, \phi^2 H^\dagger H
- \frac{1}{2} m^2 \phi^2
- \zeta\, \phi^4 \, ,
\end{align}
where $e_R$ is the right-handed electron, 
\(
L_i = (\nu_{e_i},\, e_i)_L^{\,T}
\)
are the left-handed lepton doublets with
\( e_i = e,\mu,\tau \),
and \(H\) denotes the SM Higgs doublet field.
After electroweak symmetry breaking (EWSB), the Higgs field acquires a vacuum
expectation value (VEV),
\begin{equation}
\langle H \rangle = \left( 0,\, \tfrac{v}{\sqrt{2}} \right)^T ,
\end{equation}
while the scalar field \(\phi\) does not develop a VEV, ensuring that the
\(Z_2\) symmetry remains exact.
The Higgs VEV contributes to the mass of the scalar field \(\phi\), yielding
its physical mass square
\begin{equation}
m_\phi^2 = m^2 + \lambda' v^2 .
\end{equation}
The exact \(Z_2\) symmetry prevents any mixing between \(\phi\) and the SM
Higgs boson. The only interaction connecting the dark scalar to the SM Higgs
sector arises from the quartic term \(\phi^2 H^\dagger H\), which can open an
additional Higgs decay channel \(H \to \phi \phi\) for sufficiently light
\(\phi\). However, the coupling \(\lambda'\) can be chosen to be very small,
independently of the scalar mass, thereby evading constraints from the invisible Higgs decay searches.

The mass of the vector-like fermion $\psi$ is not generated through symmetry
breaking and is entirely determined by the parameter $m_\psi$. Although
$\psi$ does not couple directly to the Standard Model fields, it communicates
with the visible sector through interactions mediated by the heavy
right-handed neutrinos $N_i$ and the scalar field $\phi$. As we shall
demonstrate in the subsequent sections, these neutrino-portal interactions
play a crucial role in linking the dynamics of the dark sector to the
generation of the matter-antimatter asymmetry of the Universe. This
interplay between dark matter and leptogenesis constitutes the central focus
of the present work.

The model addresses the origin of neutrino masses via the standard type-I
seesaw mechanism~\cite{Brdar:2019iem}, induced by the Yukawa interaction
governed by the couplings \(Y^N_{ij}\). For simplicity, we assume the Majorana
mass matrix of the right-handed neutrinos to be diagonal,
\(
m_{N_{ij}} = m_{N_i}\,\delta_{ij}
\),
so that the physical masses of the heavy neutrinos are given by \(m_{N_i}\).
The seesaw Yukawa couplings \(Y^N_{ij}\) are fixed using the Casas--Ibarra
parameterization~\cite{Casas:2001sr}.

The right-handed neutrinos $N_i$, being Majorana fermions, naturally provide
the necessary ingredients for leptogenesis through their out-of-equilibrium
decays. In the standard type-I seesaw framework, leptogenesis arises solely
from the Yukawa interactions between $N_i$, the SM leptons, and the Higgs
doublet. In the present model, this conventional picture is extended by the
additional interaction term $\bar{\psi} N_i \phi$, which directly connects
the right-handed neutrinos to the dark sector. The associated Yukawa
couplings $y_i$, being unconstrained by low-energy experiments, introduce
new sources of $C\!P$ violation beyond those present in the standard
leptogenesis scenario. As a result, the decay of $N_i$ can generate
asymmetries simultaneously in the visible and dark sectors. Moreover, this
interaction opens up a variety of scattering processes involving leptons,
dark-sector particles, and heavy neutrinos, which can act both as sources of
lepton asymmetry and as channels for transferring asymmetry between the two
sectors. In addition, when kinematically allowed, the decay of the scalar
field $\phi$ can itself generate a dark-sector asymmetry, which may
subsequently influence the evolution of the lepton asymmetry through these
scattering processes. The combined effect of these decay and scattering
channels establishes a nontrivial interplay between dark matter dynamics and
leptogenesis, which constitutes the central focus of this work.

\subsection{Collider Constraints}
Major collider constraints on the present framework arise from searches for 
heavy neutral leptons (HNLs), corresponding to the right-handed neutrinos $N_i$, at the  Large Hadron Collider. At the LHC, HNLs can be produced through their mixing with Standard Model leptons, typically via processes such as 
$pp \to W^\ast \to \ell N$, followed by $N$ decays into leptons and electroweak gauge bosons. Both of the ATLAS and the CMS collaborations have carried out extensive searches for HNLs in prompt and displaced decay channels~\cite{CMS:2022fut, ATLAS:2019kpx, ATLAS:2024rzi,
LHCb:2020wxx, ATLAS:2025-038}. These searches are sensitive to the active-sterile mixing along the masses of HNLs. For HNL masses between 100 GeV to 20 TeV, the mixing is limited to $<(0.1-1)$ \cite{ATLAS:2024rzi}. 
We shall restrict our study to heavy neutrinos with masses $m_{N_i} \gtrsim 2~\text{TeV}$ and active-sterile mixing angles smaller than $|U_{\ell N}|^2 \lesssim 1.2 \times 10^{-5}$, well below the current sensitivity of LHC searches. Within this regime, collider constraints are naturally evaded, allowing us to focus on the cosmological implications of the model.

We first analyze the framework to determine the regions of parameter space consistent with the observed dark matter relic density. Subsequently, we investigate the viability of leptogenesis within this restricted parameter space, with particular emphasis on the novel contributions arising from neutrino-portal interactions. In this setup, the interplay between the dynamics of the dark sector and lepton number generation plays a central role, enabling successful low-scale leptogenesis while simultaneously satisfying constraints on dark matter.
\section{\label{sec:dm}Dark Matter} 
Both $\phi$ and $\psi$ are odd under an exact discrete $Z_2$ symmetry, which guarantees their stability and renders them viable dark matter (DM) candidates. When kinematically allowed, the heavier of the two particles can decay into the lighter one, effectively yielding a single-component DM scenario. On the other hand, if such decays are kinematically forbidden or sufficiently suppressed, a two-component DM framework arises naturally, with both $\phi$ and $\psi$ contributing to the total dark matter relic abundance. 

As discussed in Section~\ref{sec:model}, the quartic coupling between $\phi$ and the standard Higgs field, $\lambda'$ is taken to be very small, so as to avoid collider constraints. Thus, the only interaction $\phi$ and $\psi$ have with the visible sector is through the Yukawa interaction $\bar \psi  \phi N_i$, 
controlled by the respective couplings, $y_i$. For this coupling being in the range of ${\cal O}(y_i)<10^{-10}$, we have FIMP dark matter which follows a non-thermal genesis of the dark matter relic abundance. We have explored this case and found that for a large range of dark matter masses in the GeV to TeV range, the compatible parameters lead to $y_i \sim 10^{-12}$.  However, such small couplings do not lead to any significant effect in the leptogenesis. Therefore, we shall not discuss this scenario any further in this work.

On the other hand, the WIMP scenario is of particular interest in this framework, as the presence of sizable couplings can lead to a significant impact of dark matter interactions on leptogenesis. The viable dark matter candidates in this case are dictated by the mass hierarchy among the exotic particles. Accordingly, we consider the different possibilities separately. In the following, we first demonstrate why a single-component dark matter scenario is not viable within this framework. We then systematically analyze the two-component scenarios, identify the corresponding viable regions of parameter space, and discuss their phenomenological consequences.

\subsection{Single-Component Dark Matter}
Depending on the mass hierarchy, we may have either a scalar DM scenario or a scotino (fermionic) DM scenario. In the first case of scalar DM scenario, the dark scalar, $\phi$ is lighter than the new fermion, $\psi$. The scalar DM candidate, $\phi$ will remain in thermal equilibrium with the SM particles in the early Universe through its Higgs portal interactions.  This makes it a potential WIMP candidate. The annihilation of DM will be controlled by the quartic coupling, $\lambda' \phi^2H^\dagger H$. However, the same coupling also decides the DM-nucleus scattering in the direct detection experiments. Therefore, it turned out to be quite challenging to deplete the DM sufficiently to avoid an overabundance in DMRD while satisfying the measurements at the direct detection experiments. This is a typical issue with scalar DM and Higgs portal interactions, and the present model framework is no exception.  

In the scotino DM scenario, the dark fermion $\psi$ remains in thermal equilibrium with the Standard Model (SM) particles through processes such as $\psi\,\psi \to N_i\,N_j,~\phi\,\phi$. However, for $m_\psi < (m_\phi, m_N)$, no tree-level annihilation channels are available that can efficiently deplete the $\psi$ abundance to reproduce the observed dark matter relic density during thermal freeze-out. Annihilation processes of the type $\psi\,\psi \to \text{SM}$ may arise at higher orders; however, these contributions are strongly suppressed by loop factors, the smallness of the Higgs-portal coupling $\lambda'$ required by collider constraints, and the bounds on the seesaw Yukawa couplings. 
 
In view of the above constraints, we do not pursue the single-component WIMP scenario any further in this work.

\subsection{Co-existing Fermion-Scalar Two-component Dark Matter}
In kinematic regions where $\phi$ and $\psi$ are mass-degenerate or nearly degenerate, the co-annihilation channel $\phi\,\psi \to \text{SM}$ becomes significant. In contrast, the annihilation processes $\phi\,\phi \to \text{SM}$ and $\psi\,\psi \to \text{SM}$ are highly suppressed, such that the dark matter relic abundance predominantly determined by the co-annihilation channel.

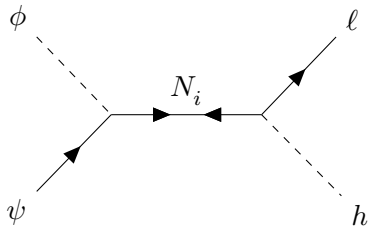
\begin{figure}[!htb]
    \centering
    \begin{tikzpicture} 
        \begin{feynman}
            \vertex (a1);
            \vertex[above left=1.4cm of a1] (a2) {\(\phi\)};
            \vertex[below left=1.4cm of a1] (a3) {\(\psi\)};
            \vertex[right=2.0cm      of a1] (a4);
            \vertex[above right=1.4cm of a4] (a5) { \( \ell \)};
            \vertex[below right=1.5cm of a4] (a6){\( h \)};
            \diagram* {
                (a2) -- [scalar] (a1),
                (a3) -- [fermion] (a1),
                (a1) -- [majorana, edge label={\( N^{}_i \)}] (a4),
                (a4) -- [fermion] (a5),
                (a4) -- [scalar] (a6)
            };
        \end{feynman}  
        \end{tikzpicture} \quad \quad \quad \quad      
\caption{Feynman diagram showing the tree-level co-annihilation process \( \phi ~ \psi \to \ell~ h \).} 
\label{Fig:two_two_scattering_analysis}
\end{figure}

The process mediated by the right-handed neutrinos, $\phi ~\psi \to \ell ~h$, is the main co-annihilation channel. 
The Feynman diagram corresponding to these processes is depicted in Fig.~\ref{Fig:two_two_scattering_analysis}. 

The temperature evolution of the yields $Y_{\phi}$ and $Y_{\psi}$ corresponding to the fields $\phi$ and $\psi$, respectively; is governed by the coupled Boltzmann equations,
\begin{align}
    H S x \frac{dY_{\phi}}{dx} &=
    \gamma^{\rm sc}\left(1-\frac{Y_{\phi}}{Y^{\rm eq}_{\phi}}
    \frac{Y_{\psi}}{Y^{\rm eq}_{\psi}}\right),~{\rm and} \nonumber \\
    H S x \frac{dY_{\psi}}{dx} &=
    \gamma^{\rm sc}\left(1-\frac{Y_{\phi}}{Y^{\rm eq}_{\phi}}
    \frac{Y_{\psi}}{Y^{\rm eq}_{\psi}}\right),
    \label{eq:Bz2}
\end{align}
where $\gamma^{\rm sc}$ denotes the thermally averaged cross section for the processes
$\phi\,\psi \to \ell\,h$, $\phi\,\psi \to \bar{\ell}\,\bar{h}$,
$\phi\,\bar{\psi} \to \ell\,h$, and $\phi\,\bar{\psi} \to \bar{\ell}\,\bar{h}$.
We have assumed that the scotino $\psi$ and its antiparticle $\bar{\psi}$ have equal number densities.

As discussed later in the context of leptogenesis, the complex Yukawa coupling $y_i$ associated with the interaction term $\bar{\psi}\,\phi\,N_i$ can generate a small asymmetry between the number densities of $\psi$ and $\bar{\psi}$. However, this effect is subdominant for the dark matter dynamics, which is primarily governed by the WIMP freeze-out process. Nevertheless, such an asymmetry in the fermionic dark sector can have important implications for leptogenesis, and this constitutes one of the key features of our framework. A detailed discussion of this aspect is presented in the leptogenesis section.

The remaining parameters in Eq.~\eqref{eq:Bz2} have their standard meanings: $x = m_\phi/T$ denotes the scaled inverse temperature, $H$ is the Hubble parameter, $S$ is the entropy density of the Universe, and $Y^{\rm eq}_{\phi,\psi}$ are the equilibrium yields. Since the equilibrium number densities of $\phi$ and $\psi$ differ primarily by their internal degrees of freedom, the ratios $Y/Y^{\rm eq}$ are approximately equal for the two species. Nevertheless, for completeness, we solve the coupled equations in Eq.~\eqref{eq:Bz2} explicitly.

While the Boltzmann equations determine the late-time comoving yields $Y_{\phi,\psi}$, cosmological observations constrain the present-day dark matter abundance in terms of the dimensionless relic density parameter $\Omega h^2$. Accordingly, the relic density contributions are expressed as
\begin{equation}
  \Omega_{\psi,\phi} h^2
  = \frac{S\,h^2\, m_{\psi,\phi}}{\rho_c}\, Y_{\psi,\phi},
  \label{eq:relicdm}
\end{equation}
where $\rho_c = \frac{3H^2}{8\pi G}$ is the critical density, with $H = 100\,h~\text{km s}^{-1}\text{Mpc}^{-1}$ and $G$ denoting Newton’s constant. The total dark matter relic abundance is then given by
\begin{equation}
  \Omega_{\rm tot} h^2
  = 2\,\Omega_{\psi} h^2 + \Omega_{\phi} h^2.
  \label{eq:relitdm}
\end{equation}
Here $\Omega_\psi h^2$ denotes the contribution from either $\psi$ or $\bar{\psi}$ separately. Since the Boltzmann equation is solved for a single scotino species with $Y_\psi = Y_{\bar{\psi}}$, the total fermionic dark matter contribution carries an explicit
factor of two. 

Notice that, for a fixed $m_{N_i}$, the relevant seesaw Yukawa couplings $Y^N_{ij}$ are fixed. Consequently, the co-annihilation cross section for the process $\phi\,\psi \to \ell\,h$ depends only on the common dark matter mass $m_\phi = m_\psi$ and the Yukawa coupling $y_i$. For a fixed relic abundance (or, equivalently, for a fixed thermally averaged effective annihilation rate) an increase in the dark matter mass must be compensated by a decrease in the coupling $y_i$, leading to an anti-correlation between $m_{\phi,\psi}$ and $y_i$. As a result, the perturbative limit $y_i = \sqrt{4\pi}$ corresponds to the minimum value of $m_{\phi,\psi}$ that reproduces the observed dark matter relic abundance.
Further, with $m_{N_1} $ in the TeV range and $m_{N_{2,3}} \gg m_{N_1}$, the effect of $N_2$ and $N_3$ mediated processes are negligible in the co-annihilation process. Our analysis indicates that the DM candidate must be in the range of ${\cal O}(10^{2})$ GeV to satisfy the observed relic density in this scenario. To remind the reader, large couplings $y_i$ required here will not influence the detection prospects at the direct detection experiments, as the process is mediated by heavy neutrinos, which do not have any direct interaction with the nucleus. 

Considering benchmark scenarios with $m_{N_1}$ in the TeV range, Table~\ref{tab:dark1} presents the lower bound on the dark matter mass corresponding to the perturbative limit of the coupling $y_1$. We keep the value of the quartic coupling, $\lambda'$ fixed to $1.0\times 10^{-10}$ all through this study. As mentioned above, this evades the darkmatter direct detection constraints, as well as the Higgs sector collider constraints. As an illustrative example, for  $m_{N_1}=2~\text{TeV}$ one finds $m_{\phi,\psi} \ge 200~\text{GeV}$.
The sensitivity of the relic abundance to variations in $m_{N_1}$ is also illustrated in the Table~\ref{tab:dark1}. A decrease in $m_{N_1}$ enhances the co-annihilation cross section for the process $\phi\,\psi \to \ell\,h$, resulting in a smaller relic density. This effect can be compensated either by reducing the $\bar{\psi}\,\phi\,N_1$ Yukawa coupling $y_1$ or by lowering the common dark matter mass $m_{\phi}=m_{\psi}$.
Conversely, increasing $m_{N_1}$ suppresses the co-annihilation cross section. To maintain the observed relic abundance in this case, the
minimum allowed value of $m_{\phi}=m_{\psi}$ shifts to larger values,
since the coupling cannot exceed the perturbative limit.
\begin{table}[!htb]
\centering
\begin{tabular}{l|c|c|c|c}
\toprule
\hline
       BPs & $m_{\phi}=m_\psi~[\text{GeV}]$  & $m_{N_1}~[\text{GeV}]$ & $y_1$ &$\Omega_{\rm tot} h^2$\\ 
       \hline \hline
BP 1  & 200 & 2000  & $2 \sqrt{\pi}$ & 0.12\\ \hline
BP 2  & 200 &  1000 & $2 \sqrt{\pi}$ & 0.04  \\ \hline
BP 3  & 200 &  3000 & $2 \sqrt{\pi}$ & 0.21 \\ \hline
BP 4  & 200 &  1000 & $\sqrt{1.56}$ & 0.12\\ \hline
BP 5  & 187 &  1000 & $2 \sqrt{\pi}$ & {0.12}\\ \hline
BP 6  & {209} &  3000 & $2 \sqrt{\pi}$ & {0.12} \\  
\hline
\bottomrule
\end{tabular}
\caption{Benchmark values of model parameters $m_{\phi}=m_{\psi}$, $m_{N_1}$, and $y_1$ and the corresponding DM relic abundance $\Omega_{tot} h^2$. The heavier right-handed neutrino masses are fixed at $m_{N_2}=10^9$ GeV and at $m_{N_3}=2.6\times 10^{13}$ GeV, and the quartic coupling $\lambda'=10^{-10}$.
~~~~~~~~~~~~~~~~~~~~~~~~~~~~~~~~~~~~~~~~~~~~~~~~~~~~~~~~~~~~~~~~~~~~~~~~~~~~~~~~~~~}
\label{tab:dark1}
\end{table}
As mentioned above, for $m_{N_1} \ll m_{N_{2,3}}$, the contributions of $N_2$ and $N_3$ to the dark matter dynamics are negligible, and their precise masses are therefore not relevant for the present discussion. However, these masses play an important role in leptogenesis, since the corresponding seesaw Yukawa couplings enter the generation of the lepton asymmetry. Accordingly, we fix
$m_{N_2} = 10^9~\text{GeV}$ and $m_{N_3} =2.6 \times 10^{13}~\text{GeV}$.

In the subsequent analysis, we fix $m_{N_1}=2~\text{TeV}$ as an illustrative benchmark point, keeping in mind the sensitivity of both the dark matter dynamics and leptogenesis, as discussed later, to this parameter space. As shown above, this choice requires a minimum dark matter mass of $m_{\phi,\psi} \simeq 200~\text{GeV}$ when the coupling saturates its perturbative limit, $y_1=\sqrt{4\pi}$. We find that such large couplings generically lead to a lepton asymmetry that is much below the observed value.

Our analysis further indicates that the generated lepton asymmetry is more sensitive to the coupling $y_1$ than to the dark matter mass, since $y_1$ enters both the $CP$ asymmetry and the relevant reaction rates. In practice, viable leptogenesis typically requires $y_1$ in the range $0.01\text{--}0.1$. Ideally, one would perform a comprehensive scan over
the dark matter mass and coupling to identify the region consistent with the relic density, and independently carry out a similar scan for leptogenesis. The overlap of these regions would then determine the parameter space compatible with both dark matter constraints and
successful leptogenesis. However, such a global scan is computationally prohibitive.

We therefore adopt a more pragmatic strategy. We fix the dark matter mass to
$m_{\phi,\psi}=250~\text{GeV}$, motivated by its proximity to the minimum mass required for the chosen neutrino-sector parameters.  The coupling $y_1$ is then scanned over a coarse range to identify regions corresponding to underabundant and overabundant dark matter. For these values of $y_1$, the resulting lepton asymmetry is subsequently computed to assess the viability of this parameter region. The results are presented in Table~\ref{tab:dark2}.
\begin{table}[!htb]
\centering
\begin{tabular}{l|c|c|c|c}
\toprule
\hline
       BPs & $m_{\phi}=m_\psi~[\text{GeV}]$  & $m_{N_1}~[\text{GeV}]$ & $y_1$ &$\Omega_{\rm tot} h^2$\\ 
       \hline \hline
BP 7  & 250 & 2000  & $0.0042$ &$0.12$\\ \hline
BP 8  & 250 & 2000  & $0.01$ & $0.032$\\ \hline
BP 9  & 250 & 2000  & 0.1 & $0.0033$\\ \hline
BP 10  & 250 & 2000  & 0.5 & $0.0028$ \\  \hline
\bottomrule
\end{tabular}
\caption{Dark matter relic abundance $\Omega_{\text{tot}} h^2$ as a function of the coupling $y_1$ for a fixed DM mass of $250~\text{GeV}$. The heavier right-handed neutrino masses are fixed at $m_{N_2}=10^9~\text{GeV}$ and $m_{N_3}=2.6\times10^{13}~\text{GeV}$.}
~~~~~~~~~~~~~~~~~~~~~~~~~~~~~~~~~~~~~~
\label{tab:dark2}
\end{table}
Let us now briefly examine the sensitivity of the above discussion to variations in $m_{N_1}$.
As shown in Table~\ref{tab:dark1}, increasing $m_{N_1}$ while keeping it within the few-TeV range modifies the minimum allowed dark matter mass by no more than $\mathcal{O}(10\%)$. For dark matter masses above these minima, one can always identify a corresponding value of the coupling $y_1$ that reproduces the observed relic abundance. We also note that an underabundant dark matter contribution is phenomenologically acceptable, provided the total dark matter density is not required to be saturated by this model alone.
\section{\label{sec:leptogenesis}Leptogenesis}
We now turn our attention to the details of leptogenesis within the framework proposed in this study, where the required lepton number violation and $C\!P$ asymmetry arise from the decay of heavy Majorana neutrinos $N_{1,2,3}$. In the conventional thermal leptogenesis scenario, such asymmetry is generated via the out-of-equilibrium decays of right-handed Majorana neutrinos into the SM leptons and the Higgs doublet. However, $C\!P$ violation arises here through the quantum corrections to the decay process, both self-energy corrections and vertex corrections. In the standard scenario, these corrections are decided by the type-1 seesaw interaction~\cite{Fukugita:1986hr,Covi:1996wh}. In this work, $y_i\bar{\psi}\phi N_i$ provides the additional self-energy and vertex correction contributions. This additional interaction modifies the dynamics of leptogenesis in several important ways. Notably, the vertex and self-energy loop corrections contributing to $C\!P$ violation in the decay processes receive new contributions from diagrams involving the fields \( \phi \) and \( \psi \). These additional contributions can potentially enhance the $C\!P$ asymmetry compared to the standard scenario, and allow a viable leptogenesis at lower energy scales. 
Notice that, apart from the standard decay channel, $N_1 \to \ell H$, the heavy neutrino $N_1$ decays to the dark sector, $N_1 \to \phi \psi$ when kinematically allowed. In this study, learning from the dark matter analysis, we restrict ourselves to a degenerate $\phi-\psi$ dark matter, which is always considerably lighter than $N_1$. The decay of heavy neutrino into dark sector particles also provides the sources of $C\!P$ asymmetry at one-loop, very similar to the standard scenario. This asymmetry, however, does not induce any lepton number asymmetry directly. At the same time, this can generate significant asymmetry in the dark sector, which in turn can influence the thermal evolution of the lepton number asymmetry. 
While strict mass degeneracy of $\psi$ with $\phi$ disallows $\phi$ decay, making it part of the dark matter spectrum, sufficient non-zero mass-splitting could allow $\phi$ decay through 3-body or 4-body final states. In principle, these 3-body and 4-body decays of $\phi$ could also generate additional $C\!P$ violation. However, in the rest of this study, we impose strict mass degeneracy of $\phi$ and $\psi$ to prohibit such $3/4$-body decays of $\phi$.
Another source of $C\!P$ violation arises from scattering processes. Scattering channels mediated by right-handed neutrinos can also violate $C\!P$ symmetry through the interference of amplitudes involving different heavy neutrinos, $N_i$ and $N_j$ with $i \neq j$. The relevant scattering processes in this case are
$\phi\,\psi \to \ell H$, $\phi^{\dagger}\,\bar{\psi} \to \ell H$,
$\ell\,\phi \to H\,\psi$, $\bar{\ell}\,\phi \to H^{\dagger}\,\psi$,
$\bar{\ell}\,\psi \to H^{\dagger}\,\phi$, and $\ell\,\psi \to H\,\phi$.

In the standard leptogenesis, scattering processes typically contribute to washout effects, often significantly reducing the lepton asymmetry generated by $C\!P$-violating decays. In the present framework, however, scattering processes play three distinct roles: they act as source terms that generate lepton asymmetry through their intrinsic $C\!P$ violation, as washout terms that suppress the generated asymmetry, and as asymmetry-transfer processes that redistribute the asymmetry between the dark sector and the leptonic sector, and vice versa.

Before proceeding to the thermal evolution of the yields of the various particle species, we first analyze the $C\!P$ asymmetries arising in the different processes.
\subsection{\texorpdfstring{$C\!P$ Asymmetry}{CP Asymmetry} \label{sec:CPasymmetry}}
\label{sec:cp asymmetry}
The $C\!P$ asymmetry arising from the decay of right-handed neutrinos,  as discussed in Refs.~\cite{Fukugita:1986hr,Yoshimura:1978ex,Buchmuller:2004nz}, originates from the interference between tree-level and loop-level diagrams. In the standard case, the loop-level contributions to the decay of right-handed neutrinos arise from diagrams with leptons and the Higgs boson running in the loop. In contrast, within the framework considered here, the $C\!P$ asymmetry in the decay of $N_1$ receives contributions not only from the standard lepton--Higgs loops but also from additional quantum corrections
involving the dark-sector fields $\psi$ and $\phi$.
While the general idea of linking dark sector dynamics with leptogenesis has been explored previously, notably in Ref.~\cite{Falkowski:2011xh}, the scenario studied here is substantially different. Our framework does not rely on asymmetric dark matter, but instead considers a multi-component WIMP scenario in which the dark matter abundance is set by thermal freeze-out. Furthermore, scattering processes play a qualitatively new role in our analysis, contributing not only to washout but also directly to the generation and transfer of lepton asymmetry, hence an important distinction of our study in comparison to Ref.~\cite{Falkowski:2011xh}. We discuss the various contributions to the $C\!P$ asymmetry in the following.
\subsubsection{\texorpdfstring{$C\!P$ Asymmetry from the decay of $N_1 \to \ell H$}{CP Asymmetry from the decay of N1 -> l H}}
\label{sec:cpn1}
\begin{figure}[!htb] 
    \centering
	\begin{tikzpicture}  
	\begin{feynman}
	\vertex (a1){\(N_1\)};
	\vertex[right=1.4cm of a1] (a2) ;
	\vertex[right=1.4cm of a2] (a3) ;
	\vertex[right=1.1cm of a3] (a4) ;
	\vertex[above right=1cm of a4] (a5){\( H  \)};
	\vertex[below right=1cm of a4] (a6){\( \ell \)} ;
	\diagram* { (a2) --  [majorana](a1),(a4)  --[scalar ] (a5),(a4) --[fermion] (a6), (a3) --[majorana,edge label= {\( N_j \)}] (a4),  (a2) --[ scalar,half left, edge label= {\( H/\phi \)}] (a3),(a2) --[ fermion,half  right, edge label= {\( \ell/ \psi \)} ] (a3),    
	};
	\end{feynman}  
 \end{tikzpicture} \hskip 20mm
 \begin{tikzpicture}
      \quad \quad
      \begin{feynman}
	\vertex (b1){\(N_1\)};
	\vertex[right=1.4cm of b1] (b2) ;
	\vertex[above right=1.4cm of b2] (b3) ;
	\vertex[below right=1.4cm of b2] (b4) ;
	\vertex[ right=1cm of b4] (b5){\( H \)};
	\vertex[ right=1cm of b3] (b6) {\( \ell \)} ;
	\diagram* { (b2) --  [majorana](b1),(b4)  --[scalar ] (b5),(b3) --[fermion] (b6), (b3) --[majorana,edge label= {\( N_j \)}] (b4),  (b2) --[ scalar, edge label= {\( H   \)}] (b3),(b4) --[anti fermion, edge label= {\( \ell \)} ] (b2),    
	};
	\end{feynman} 
	\end{tikzpicture}  
 \quad
\caption{Feynman diagrams representing the one-loop corrections to the decay \(N_1 \to \ell H\) including the dark sector contributions to the self-energy.} 
\label{Fig:CP-Asymmetry_one_loop}
\end{figure}
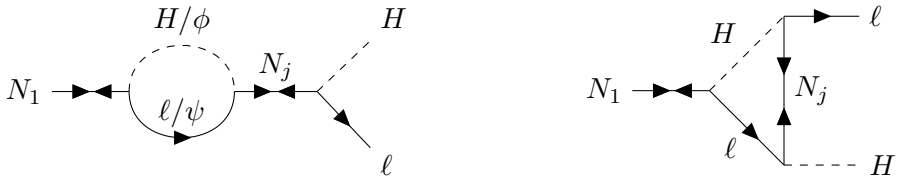

The  $C\!P$ asymmetry arising from the decay of $N_1$ is defined as 
\begin{align}
    \epsilon_1  = \frac{\Gamma(N_1 \to \ell ~H) - \Gamma(N_1 \to \bar{\ell} H^{\dagger})}{\Gamma(N_1 \to \ell H) + \Gamma(N_1 \to \bar{\ell} H^{\dagger})} ,
\end{align}
where $\Gamma(N_1 \to \ell ~H)$, and $\Gamma(N_1 \to \bar{\ell} ~H^{\dagger})$ are the decay widths of $N_1$ into leptons and antileptons, respectively. The quantum corrections to the decay process at one loop level leading to $C\!P$ violation are shown in Fig.~\ref{Fig:CP-Asymmetry_one_loop}. Notice that, in addition to the standard contributions arising through the presence of $H$ and $\ell$ in the loop, in the present case, the quantum corrections get contributions from the interaction of $N_i$ with the dark sector particles, $\phi$ and $\psi$. These additional contributions are proportional to the new Yukawa couplings, $y_i$. While in the DM dynamics, only the magnitudes of these couplings $|y_i|$ play a role; the complex nature of the couplings $y_i$ is vital for Leptogenesis.
The $C\!P$ asymmetry including the contributions arising from the dark sector dynamics is given  below with total $C\!P$ asymmetry $\epsilon_1$ getting contributions from the self-energy ($\epsilon^{s}$) and vertex correction ($\epsilon^{v}$):
 \begin{align}
 \epsilon_1 & = \epsilon^s + \epsilon^v, \label{eq:epsilon1} ~{\rm where}\\
\epsilon^{s} &= \frac{1}{8 \pi K_{11}} \sum_{j=2}^{3} \frac{m_{N_1}}{m_{N_1}^2 - m_{N_j}^2} 
\operatorname{Im} \left( m_{N_j} K_{1j}^2 + m_{N_1} \kappa_{j1} K_{1j} + m_{N_j} \kappa_{1j} K_{1j} \right) , ~{\rm and } 
\label{eq:self} \\
\epsilon^{v} &= \frac{1}{8 \pi K_{11}} \sum_{j=2}^{3} \operatorname{Im}\left(K_{1j}^2\right) \mathcal{F}\left(\frac{m_{N_j}^2}{m_{N_1}^2}\right),\label{eq:vertex}
\end{align}  
$K_{ij} = (Y^{N\dagger} Y^N)_{ij}$, and $\kappa_{ij} =(y_i^\dagger y_j)$ respectively. The function \[\mathcal{F}(x) = \sqrt{x}\left[1 + (1 + x) \ln \left(\frac{x}{1 + x}\right)\right]\] 
is related to vertex loop factor. In the self-energy correction contribution,  $\epsilon^s$ given in  Eq. \ref{eq:self} has three different terms. The first term is related to the standard self-energy part of $C\!P$ asymmetry, whereas the second and the third terms are the novel contributions due to $\phi$ and $\psi$ in the loop. These additional terms are crucial for low-scale leptogenesis, without requiring mass degeneracy of right handed neutrinos. 
\subsubsection{\texorpdfstring{$C\!P$ Asymmetry from decay of $N_1 \to \phi \psi$}{CP Asymmetry from decay of N1 -> phi psi}}
\label{sec:cp_asymmetry_from_ntophipsi}
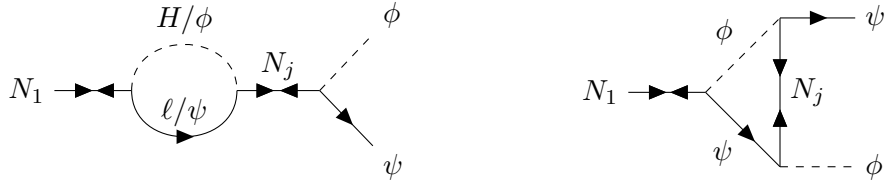
\begin{figure}[!htb] 
    \centering
\begin{tikzpicture}  
	\begin{feynman}
	\vertex (a1){\(N_1\)};
	\vertex[right=1.4cm of a1] (a2) ;
	\vertex[right=1.4cm of a2] (a3) ;
	\vertex[right=1.1cm of a3] (a4) ;
	\vertex[above right=1cm of a4] (a5){\( \phi  \)};
	\vertex[below right=1cm of a4] (a6){\( \psi \)} ;
	\diagram* { (a2) --  [majorana](a1),(a4)  --[scalar ] (a5),(a4) --[fermion] (a6), (a3) --[majorana,edge label= {\( N_j \)}] (a4),  (a2) --[ scalar,half left, edge label= {\( H /\phi \)}] (a3),(a2) --[ fermion,half  right, edge label= {\( \ell/ \psi \)} ] (a3),    
	};
	\end{feynman}  
 \end{tikzpicture} \hskip 20mm
 \begin{tikzpicture}
      \quad \quad
      \begin{feynman}
	\vertex (b1){\(N_1\)};
	\vertex[right=1.4cm of b1] (b2) ;
	\vertex[above right=1.4cm of b2] (b3) ;
	\vertex[below right=1.4cm of b2] (b4) ;
	\vertex[ right=1cm of b4] (b5){\( \phi \)};
	\vertex[ right=1cm of b3] (b6) {\( \psi \)} ;
	\diagram* { (b2) --  [majorana](b1),(b4)  --[scalar ] (b5),(b3) --[fermion] (b6), (b3) --[majorana,edge label= {\( N_j \)}] (b4),  (b2) --[ scalar, edge label= {\(  \phi \)}] (b3),(b4) --[anti fermion, edge label= {\( \psi \)} ] (b2),    };
	\end{feynman} 
	\end{tikzpicture}  
 \quad
\caption{Feynman diagrams representing the one-loop corrections to the decay \(N_1 \to \phi ~\psi\).}
\label{Fig:CP-Asymmetry_one_loop_DM1}
\end{figure}

As discussed before in Subsubsection~\ref{sec:cpn1}, 
the $C\!P$ asymmetry generated by the decay of $N_1$ into the visible sector is influenced by its interaction with the dark sector. 
At the same time, the decay of $N_1$ into the dark sector also generates the $C\!P$ asymmetry, leading to asymmetric generation of $\psi$ and its antiparticle, $\bar \psi$. The $C\!P$ asymmetry in this case is defined as 
\begin{equation*}
    \epsilon_{DM} =\frac{\Gamma(N_1 \to \phi ~\psi)-\Gamma(N_1 \to {\phi} ~ \bar{\psi})}{\Gamma(N_1 \to \phi ~\psi)+\Gamma(N_1 \to {\phi}~ \bar{\psi})},
\end{equation*}
where $\phi$ is its own antiparticle.
Feynman diagrams associated with such $C\!P$ asymmetry sources are shown in Fig.~\ref{Fig:CP-Asymmetry_one_loop_DM1}. Here, the loop level decay of $N_1 \to \ell H$ provides the additional complex phase. This additional phase gives an additional contribution to the $\epsilon_{DM}$. It raises a very interesting question: Does dark sector asymmetry affect the visible sector (lepton-antilepton) asymmetry? We will address it in the numerical discussion in section~\ref{sec:numericalanalysis} where the dark sector asymmetry feeding into the lepton-antilepton asymmetry has a significant role in certain regions of the parameter space.
The $C\!P$ asymmetry arising in the decay $N_1\to \phi\psi$ is given by
\begin{align}
\epsilon_{DM}&= \epsilon^{s}_{DM} + \epsilon^{v}_{DM},\label{eq:epsilonDM} ~{\rm where} 
\\
\epsilon^{s}_{DM} &= \frac{1}{8 \pi ~\kappa_{11}} \sum_{j=2}^{3} \frac{m_{N_1}}{m_{N_1}^2 - m_{N_j}^2} 
\operatorname{Im} \left( m_{N_j} \kappa_{1j}^2 + m_{N_1} K_{j1} \kappa_{1j} + m_{N_j} K_{1j} \kappa_{1j} \right), \label{eq:selfDM} ~{\rm and}  \\
\epsilon^{v}_{DM} &= \frac{1}{8 \pi ~\kappa_{11}} \sum_{j=2}^{3} \operatorname{Im}\left(\kappa_{1j}^2\right) \mathcal{F}\left(\frac{m_{N_j}^2}{m_{N_1}^2}\right), 
\label{eq:cpDDM}
\end{align}  
respectively.
\subsubsection{\texorpdfstring{$C\!P$ Asymmetry from $2\to2$ Scattering process  $\phi \psi \to \ell H$}{CP Asymmetry from 2->2 Scattering process phi psi -> l H}}
\label{sec:cp_asymmetry_from_scattering_phipsilh}

\begin{figure}[!htb]
    \centering
\begin{tikzpicture} 
	\begin{feynman}
	\vertex (a1);
	\vertex[below left=1.4cm of a1] (a2){\(\phi\)} ;
	\vertex[above left=1.4cm of a1] (a3){\( \psi  \) } ;
	\vertex[right=15mm of a1] (a4) ;
	\vertex[above right=1.4cm of a4] (a5);
	\vertex[below right=1.4cm of a4] (a6); 
    \diagram* { (a2) --  [scalar](a1), (a3) --  [fermion](a1)   
	};
	\vertex (a11);
	\vertex[right=1.4cm of a11] (a12) ;
	\vertex[right=1.4cm of a12] (a13) ;
	\vertex[right=8mm of a13] (a14) ;
	\vertex[above right=1.2cm of a14] (a15){\( H  \)};
	\vertex[below right=1.2cm of a14] (a16){\( \ell \)} ;
    \diagram* { (a11) --  [majorana, edge label ={\( N_{i/j}\)}](a14),(a14)  --[scalar ] (a15),(a14) --[fermion] (a16),     };
	\end{feynman}  
	\end{tikzpicture} 
 \caption{Feynman diagrams of the scattering process \( \phi ~\psi \to \ell ~ H \) ($i \neq j $).}
\label{Fig:Cp-Asymmetry_from_scattering1}
\end{figure}
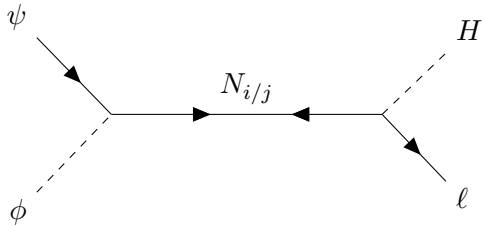

A new contribution to the total lepton asymmetry comes from the scattering process, \(\psi~\phi\to H~\ell\) mediated by heavy neutrinos $N_{1,2,3}$. The corresponding Feynman diagram is shown in Fig.~\ref{Fig:Cp-Asymmetry_from_scattering1}. The importance of these $C\!P$ asymmetry sources is that they affect the dark sector as well as visible sector asymmetries both. Therefore, these channels turn into a bridge between the dark sector and the visible sector asymmetries. An asymmetry generated in the dark sector is transferred to the visible sector and vica-versa.  
The $C\!P$ asymmetry generated by scattering process $\psi~ \phi \to H ~\ell $ is parameterized as ~\cite{PhysRevC.77.065501}
\begin{align}
    \epsilon_2 &= \frac{\sigma(\psi~\phi\to H~\ell) - \sigma(\bar{\psi}~\phi\to {H}^\dagger ~\bar{\ell})}{\sigma(\psi ~\phi\to H~\ell) + \sigma(\bar{\psi} ~\phi\to {H}^\dagger ~\bar{\ell})},~   
    \label{eq:cpep2}
\end{align}
where $\sigma$'s denote the cross sections of corresponding processes. At the tree level, interference between two different scattering \(\ \phi~\psi \to \ell ~H \) processes that are mediated by different heavy right-handed neutrinos provides the additional sources of $C\!P$ asymmetry. This phenomenon illustrates how multiple pathways in particle scattering can lead to significant effects in terms of $C\!P$ violation. Here, we take an approximation where $m_{N_1}\ll m_{N_2} \ll m_{N_3}$. In this scenario, the dominant contribution to the denominator arises from the process that is mediated by the lightest heavy right-handed neutrino \( N_1 \), whereas $\epsilon_2$ itself gets contributions from the interference between $N_1$ and $N_2$ mediated channels. In the center of mass frame of the process (with the center of mass energy $\sqrt{s}$), the generated asymmetry is given by
\begin{align}
    \epsilon_2 =& \frac{ 4 ~{\rm Im}[~y_1^* y_2 ~Y_{1 \alpha}^N (Y_{2 \alpha}^N)^* ]~ {\rm Im}[ D^{*}_1 D_2 ] }{   |y_1|^2 ~ K_{11} ~  |D_2|^2}, 
    \label{eq:ep2b}
\end{align}
where $K_{11} = (Y^{N\dagger} Y^N)_{11}$. The parameters $y_{1,2}$ and $Y^N_{(1,2)\alpha}$, with $\alpha = 1,2,3$, denote the Yukawa couplings of $N_1$ to the dark sector and to the visible sector, respectively. Besides that $D_i = s-m^2_{N_i} + i ~m_{N_i} \Gamma_i$, with $i=1,2$, denotes the denominators associated with the propagators of different right-handed neutrinos with $\Gamma_i$ being the total decay width of the mediating heavy neutrino $N_i$. After substituting $ D_i$ into Eq.~\ref{eq:ep2b}, \( \epsilon_2 \) becomes: 
\begin{align}
       \epsilon_2 =& \frac{ 4 ~{\rm Im}[y_1^* y_2 ~Y_{1 \alpha}^N (Y_{2 \alpha}^N)^* ] \left[ (s - m_{N_1}^2)~m_{N_2} ~\Gamma_2  -(s - m_{N_2}^2)~m_{N_1} ~\Gamma_1  \right] }{   |y_1|^2 ~ K_{11}  ~ \left[ (s-m_{N_2})^2+ m^2_{N_2}~ \Gamma^2_2  \right]}.
       \label{eq:ep2c}
\end{align}It is clear from Eq.~\ref{eq:ep2c} that $\epsilon_2$ depends on the center of mass energy $\sqrt{s}$, the Yukawa couplings $y$ and $Y$, the masses of right-handed neutrinos, $m_{N_i}$, and the total decay widths of right-handed neutrinos. Here, for $m_{N_3}$ being very heavy in comparison to $m_{N_1}$ and $m_{N_2}$, the contribution from $N_3$ is highly suppressed, and therefore has been neglected in our analysis.


\subsubsection{\texorpdfstring{$C\!P$ Asymmetry from $2\to2$ Scattering process  $\phi^{\dagger} \bar \psi \to \ell H$}{CP Asymmetry from 2->2 Scattering process phi-bar psi-bar -> l H}}

\label{sec:cp_asymmetry_from_scattering_barphibarpsilh}
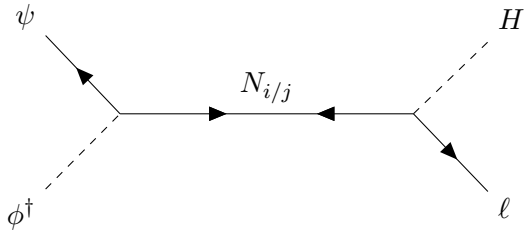
\begin{figure}[!htb]
    \centering
\begin{tikzpicture}[rotate=180] 
	\begin{feynman}
	\vertex (a1);
	\vertex[below left=1.4cm of a1] (a2){\(\phi^{\dagger} \)} ;
	\vertex[above left=1.4cm of a1] (a3){\( \bar\psi  \) } ;
	\vertex[right=2.0cm of a1] (a4) ;
	\vertex[above right=1.4cm of a4] (a5);
	\vertex[below right=1.4cm of a4] (a6); 
    \diagram* { (a2) --  [scalar](a1), (a1) --  [fermion](a3)   
	}; 
	\vertex (a11);
	\vertex[right=1.4cm of a11] (a12) ;
	\vertex[right=1.4cm of a12] (a13) ;
	\vertex[right=1.1cm of a13] (a14) ;
	\vertex[above right=1.4cm of a14] (a15){\( H  \)};
	\vertex[below right=1.4cm of a14] (a16){\( \ell \)} ;
    \diagram* { (a11) --  [majorana, edge label ={\( N_{i/j}\)}](a14),(a14)  --[scalar ] (a15),(a14) --[fermion] (a16),     };
	\end{feynman}  
	\end{tikzpicture} 
 \caption{Feynman diagrams for the scattering process \( \phi^\dagger ~\bar \psi \to \ell ~ H \)($i \neq j$).}
\label{Fig:Cp-Asymmetry_from_scattering2}
\end{figure}

Another scattering process $ \phi ~\bar \psi   \to \ell ~H $ also violates the $C\!P$ asymmetry. The relevant  Feynman diagram is shown in Fig.~\ref{Fig:Cp-Asymmetry_from_scattering2}. Similar to the scattering process \(\ \phi~\psi \to \ell ~H \), this channel also works as a bridge between the dark sector and the visible sector asymmetries. 
The $C\!P$ asymmetry generated by the scattering process $\phi^{\dagger} ~\bar \psi   \to \ell ~ H$ is parameterized as 
\begin{align}
    \epsilon'_2 &= \frac{\sigma(\bar{\psi}~\phi\to H~\ell) - {\sigma}( \psi~\phi\to H^{\dagger}~ \bar{\ell})}{\sigma(\bar{\psi}~\phi\to H~\ell) + {\sigma}( \psi~\phi\to H^{\dagger}~\bar{\ell})},
    \label{eq:ep2bar}
\end{align}
where $\sigma$ denotes the cross section of the associated process. The $C\!P$ asymmetry associated with this process $\epsilon'_2$ originates from the interference of two different $N_i$ mediated processes shown in Fig.~\ref{Fig:Cp-Asymmetry_from_scattering2}.  For being insignificant, the contributions from $N_3$ has been ignored in this analysis. After putting the cross-sections of relevant processes in Eq. \ref{eq:ep2bar}, $\epsilon'_2$ in
the center of mass frame of the process becomes,
\begin{align}
    \epsilon'_2 = \frac{ 4~ {\rm Im}[y_1 y_2^*~ Y_{1 \alpha}^N (Y_{2 \alpha}^N)^* ]~ {\rm Im}[ D^{*}_1 D_2 ] }{   |y_1|^2 ~ K_{11} ~  |D_2|^2}, 
    \label{eq:ep2pb}
\end{align} 
where $\alpha=1,2,3$ is summed over, and $D_i$ denotes the denominator of the corresponding propagator. Substituting the explicit expression for this $D_i$ factor, $\epsilon'_2$ is given by 
\begin{align}
     \epsilon'_2 =& \frac{ 4 ~{\rm Im}[y_1 y_2^*~ Y_{1 \alpha}^N (Y_{2 \alpha}^N)^* ] \left[ (s - m_{N_1}^2)~m_2~ \Gamma_2  -(s - m_{N_2}^2)~m_{N_1} ~\Gamma_1  \right] }{   |y_1|^2 ~ K_{11}  ~ \left[ (s-m_{N_2})^2+ m^2_{N_2}~ \Gamma^2_2  \right]}.
     \label{eq:ep2pb2}
\end{align}
Notice that, as the total decay width of $N_1$ depends on the mass spectrum of the dark sector, therefore $\epsilon_2$ and $\epsilon'_2$ indirectly depend on the masses of DM candidates. 
\subsubsection{\texorpdfstring{$C\!P$ Asymmetry from $2\to2$ Scattering t-channels}{CP Asymmetry from 2->2 Scattering t-channels}}
\label{sec:cp_asymmetry_from_tscattering_barphibarpsilh}

\begin{figure}[!htb]
    \centering
\begin{tikzpicture} 
	\begin{feynman}
	\vertex (a1);
	\vertex[left=1.2cm of a1] (a2){\( \ell \)} ;
	\vertex[below =1.4cm of a1] (a3);
	\vertex[left =1.2cm of a3] (a4){\(\phi \)} ;
    \diagram* {(a2)--[fermion](a1), (a1) --  [majorana, edge label ={\( N_{i/j}\)}](a3), (a3) --  [scalar](a4)   
	};
	\vertex (a11);
	\vertex[right=1.2cm of a11] (a12){\(H \)} ;
	\vertex[below=1.4cm of a11] (a13) ;
	\vertex[right=1.2cm of a13] (a14){\(\psi \)} ;
    \diagram* { (a11)  --[scalar ] (a12),(a13)  --[fermion ] (a14)     };
	\end{feynman}  
	\end{tikzpicture} \quad 
    \begin{tikzpicture} 
	\begin{feynman}
	\vertex (a1);
	\vertex[left=1.2cm of a1] (a2){\( \bar \ell \)} ;
	\vertex[below =1.4cm of a1] (a3);
	\vertex[left =1.2cm of a3] (a4){\(\phi \)} ;
    \diagram* {(a1)--[fermion](a2), (a1) --  [majorana, edge label ={\( N_{i/j}\)}](a3), (a3) --  [scalar](a4)   
	};
	\vertex (a11);
	\vertex[right=1.2cm of a11] (a12){\(  H^\dagger \)} ;
	\vertex[below=1.4cm of a11] (a13) ;
	\vertex[right=1.2cm of a13] (a14){\(\psi \)} ;
    \diagram* { (a11)  --[scalar ] (a12),(a13)  --[ fermion ] (a14)     };
	\end{feynman}  
	\end{tikzpicture}
  \begin{tikzpicture} 
	\begin{feynman}
	\vertex (a1);
	\vertex[left=1.2cm of a1] (a2){\( \bar \ell \)} ;
	\vertex[below =1.4cm of a1] (a3);
	\vertex[left =1.2cm of a3] (a4){\(\psi \)} ;
    \diagram* {(a1)--[fermion](a2), (a1) --  [majorana, edge label ={\( N_{i/j}\)}](a3), (a4) --  [fermion](a3)   
	};
	\vertex (a11);
	\vertex[right=1.2cm of a11] (a12){\(H^\dagger \)} ;
	\vertex[below=1.4cm of a11] (a13) ;
	\vertex[right=1.2cm of a13] (a14){\(\phi \)} ;
    \diagram* { (a11)  --[scalar ] (a12),(a13)  --[scalar ] (a14)     };
	\end{feynman}  
	\end{tikzpicture}  \quad 
    \begin{tikzpicture} 
	\begin{feynman}
	\vertex (a1);
	\vertex[left=1.20cm of a1] (a2){\( \ell \)} ;
	\vertex[below =1.4cm of a1] (a3);
	\vertex[left =1.20cm of a3] (a4){\(\psi \)} ;
    \diagram* {(a2)--[fermion](a1), (a1) --  [majorana, edge label ={\( N_{i/j}\)}](a3), (a4) --  [fermion ](a3)   
	};
	\vertex (a11);
	\vertex[right=1.2cm of a11] (a12){\(H \)} ;
	\vertex[below=1.4cm of a11] (a13) ;
	\vertex[right=1.2cm of a13] (a14){\(\phi \)} ;
    \diagram* { (a11)  --[scalar ] (a12),(a13)  --[scalar ] (a14)     };
	\end{feynman}  
	\end{tikzpicture}
\caption{Feynman diagrams for the scattering process \( \phi~\ell   \to \psi~ H \), \( \phi~\bar \ell   \to \psi~ H^\dagger\), \( \psi~\bar \ell   \to \phi~ H^\dagger \) and \( \psi~ \ell   \to \phi~ H \), where $i \neq j$. 
}
\label{Fig:Cp-Asymmetry_from_scatteringt1}
\end{figure}
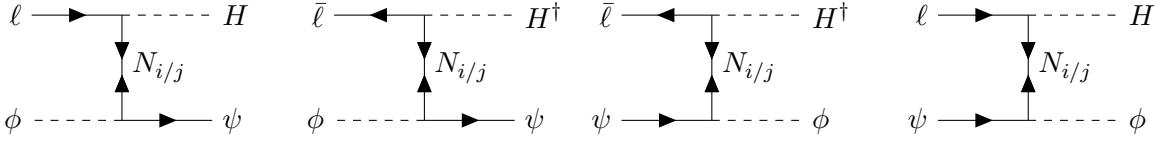

Apart from the above $C\!P$ asymmetry generating $2\to2$  s-channel scattering processes, there are several $2\to2$ t-channel scattering  processes that violate $C\!P$ symmetry. The relevant Feynman diagrams are listed in Fig.~\ref{Fig:Cp-Asymmetry_from_scatteringt1}. Notice that both $\phi$ and $\psi$ are DM candidates. Certain t-channel scattering processes allow the transfer of asymmetry generated in the dark sector to the visible sector and vice-versa, whereas in a few other t-channel processes, this asymmetry transfer does not occur.
The $C\!P$ asymmetry parameter generated by scattering processes ${\ell ~\phi \xrightarrow{t} H~ \psi}$, ${\bar{\ell} ~\phi \xrightarrow{t} H^\dagger~ \psi}$, ${\bar{ \ell}~ \psi \xrightarrow{t} H^{\dagger}~ \phi}$ and ${\ell ~\psi \xrightarrow{t} H ~\phi}$ are given by
\begin{align}
    \epsilon^t_{1} &=\frac{\Gamma(\ell ~\phi \xrightarrow{t} H ~\psi)-\Gamma( \bar{\ell}  ~\phi \xrightarrow{t} H^\dagger~ \bar{ \psi})}{ \Gamma(\ell~ \phi \xrightarrow{t} H ~ \psi)+\Gamma( \bar{\ell}  ~\phi \xrightarrow{t} H^\dagger~ \bar{\psi})} =  - \frac{\Gamma(\bar{\ell} ~\psi \xrightarrow{t} H^{\dagger} ~\phi)-\Gamma(  \ell ~ \bar{\psi} \xrightarrow{t}  H  ~\phi)}{\Gamma(\bar{ \ell} ~\psi \xrightarrow{t} H^{\dagger} ~\phi) + \Gamma(  \ell ~ \bar{\psi} \xrightarrow{t}  H ~ \phi)}, \\
\epsilon^t_{2}  &= \frac{\Gamma(\bar{\ell} ~\phi \xrightarrow{t} H^{\dagger} ~\psi)-\Gamma(\ell ~ \phi \xrightarrow{t}  H ~\bar{\psi})}{ \Gamma(\bar{\ell} ~\phi \xrightarrow{t} H^{\dagger} ~\psi)+\Gamma(  \ell ~ \phi \xrightarrow{t}  H ~\bar{\psi})} = - \frac{\Gamma(\ell ~\psi \xrightarrow{t} H~\phi)-\Gamma( \bar{\ell} ~ \bar{\psi} \xrightarrow{t} H^{\dagger}  ~\phi)}{\Gamma(\ell ~\psi \xrightarrow{t} H ~\phi) + \Gamma( \bar{\ell} ~ \bar{\psi} \xrightarrow{t} H^{\dagger} ~ \phi}, 
\end{align}
respectively. The $C\!P$ asymmetry parameters $\epsilon^{t}_1$ and $\epsilon^{t}_2$ in the center of mass frame of the process are given as follows
\begin{align}
  \epsilon^{t}_1 &=  \frac{4 ~{\rm Im}[y^{*}_1 y_2 ~Y^N_{1\alpha} (Y^N_{2\alpha})^{*}]~{\rm Im}[D^{t^*}_1D^{t}_2]}{ |y_1|^2~K_{11} ~|D^{t}_2|^2}, ~{\rm and} \label{eq:cpscdmt1}\\
 \epsilon^{t}_2 &=  \frac{4~{\rm Im}[y^{*}_1 y_2 ~(Y^N_{1\alpha})^{*} Y^N_{2\alpha}]~{\rm Im}[D^{t^*}_1D^{t}_2]}{ |y_1|^2~K_{11} ~|D^{t}_2|^2}, 
   \label{eq:cpscdmt2}
\end{align} 
where $y_{1,2}$ and $Y^N_{(1,2)\alpha}$ with $\alpha=1,2,3$ are Yukawa couplings. 
With the propagator factor,  $D^{t}_i = t-m^2_{N_i} + i ~m_{N_i} \Gamma_i$, the expressions for $\epsilon^t_{1}$ and $\epsilon^t_{2}$ are given by
\begin{align}
\epsilon^{t}_1 &= \frac{4~\mathrm{Im}\left[y^{*}_1 y_2 ~Y^N_{1\alpha} (Y^N_{2\alpha})^{*} \right] \left[ (t - m_{N_1}^2)~m_{N_2}~\Gamma_2 - (t - m_{N_2}^2)~m_{N_1}~\Gamma_1 \right]}{ |y_1|^2~K_{11} \left[ (t - m_{N_1}^2)^2 + m_{N_1}^2~\Gamma_1^2 \right]},~{\rm and} \label{eq:cpscdmmt1} \\[10pt]
\epsilon^{t}_2 &= \frac{4~\mathrm{Im}\left[y^{*}_1 y_2 ~(Y^N_{1\alpha})^{*} Y^N_{2\alpha}\right] \left[ (t - m_{N_1}^2)~m_{N_2}~\Gamma_2 - (t - m_{N_2}^2)~m_{N_1}~\Gamma_1 \right]}{ |y_1|^2~K_{11} \left[ (t - m_{N_2}^2)^2 + m_{N_2}^2~\Gamma_2^2 \right]}, \label{eq:cpscdmmt2}
\end{align}
respectively, where $m_{N_i}$ are the right handed neutrino masses of $N_i$, and $\Gamma_i$ are the total decay widths of $N_i$. The imaginary parts of coupling products and denominator of propagator products are responsible for generating the $C\!P$ violating phase. It is clear that the complex couplings, mass splittings, and decay widths of intermediate particles play a crucial role in generating $C\!P$ violating phases. The dependency of $\epsilon^{t}_1$ on right-handed neutrino masses and couplings is the same as that in the case of  $\epsilon_2$. 

In total, there are ten $C\!P$ asymmetry sources in this framework. Two are associated with the decays of $N_1$. Another six $C\!P$ asymmetry sources are connected to the scattering processes (two from s-channel processes and rest four from t-channel 
processes). The last two $C\!P$ asymmetry sources associated with three-body decay and four-body decay of $\phi$. These last two sources of $C\!P$ asymmetry are not relevant in our analysis due to mass degeneracy of $m_{\phi}$ and $m_{\psi}$. 

\subsection{Time Evolution of the Number Densities of Particles} 
\label{subsec:boltzmanneq}
The time evolution of the number densities of $N_1$, $\phi$, $\psi$, and $\bar{\psi}$ is crucial for studying the asymmetries in both the dark and the visible sectors. The evolution of the particle number densities when the system is out of thermal equilibrium is governed by the Boltzmann equations. In scenarios where the masses of the RHNs $N_i$ satisfy the hierarchy $m_{N_3} \gg m_{N_2} \gg m_{N_1}$, any lepton asymmetry generated by the decays of the heavier neutrinos $N_3$ and $N_2$ at high temperatures is efficiently washed out before the EWSB. As a result, the lepton asymmetry that survives to lower temperatures is predominantly produced by the decay of the lightest right-handed neutrinos, $N_1$. The evolution of the number density $N_1$ is influenced by several processes, including its decay and scattering processes. Similarly, the number density of the leptons is decided by processes involving their production (like $N_1$ decay and scattering processes), and annihilation (like the inverse decay and scattering processes). The processes relevant to the visible sector asymmetry and the dark sector asymmetry in our case are presented in Appendix~\ref{decayandscatteringprocess}. These interactions collectively determine the abundance of $N_1$, $\phi$, $\psi$, and the resulting lepton asymmetry. The generated lepton asymmetry can subsequently be changed into the observed baryon asymmetry of the Universe via sphaleron processes~\cite{Aoyama:1987nd}. The relevant Boltzmann equations for yields $Y_{N_1}=\frac{n_{N_1}}{s},$ $Y_L=\frac{n_l-n_{\bar{l}}}{s}$ and $Y_{\Delta \psi}=\frac{n_{\psi}-n_{\bar{\psi}}}{s}$ are as follows:
\begin{align}
    H S z ~\frac{dY_{N_1}}{dz} &= \left( 1 - \frac{Y_{N_1}}{Y_{N_1}^{\text{eq}}} \right) 
     \left( \gamma_{D_1} + \gamma_{D_2} + 2 \gamma_s + 2\gamma_{t_1}+ 2\gamma_{t_2} +2 \gamma_V \right) \label{eq:bz1}, \\
    H S z ~\frac{dY_{L}}{dz} &= \left( \frac{Y_{N_1}}{Y^{\text{eq}}_{N_1}} - 1 \right) \epsilon_1 \gamma_{D1} 
    + {2\Theta^{{\phi \psi}}_{\epsilon_2}}   + {2\Theta^{{\phi^{\dagger} \bar \psi}}_{\epsilon'_2} } - {2 \Theta^{{\phi \ell}}_{\epsilon^{t}_1}}  - {2 \Theta^{{\bar \ell \psi }}_{\epsilon^{t}_1}}  +{2 \Theta^{{\phi \bar \ell  }}_{\epsilon^{t}_2}}  +{2 \Theta^{{ \ell \psi }}_{\epsilon^{t}_2}} \nonumber \\
    + & \frac{Y_{\Delta \psi}  }{Y^{eq}_{\psi}} \left( \frac{\gamma_{\phi \psi}}{2}  + \frac{\gamma'_{\phi \psi}}{2}   + \frac{\gamma^{t}_{\phi \ell}}{2}   -  \frac{\gamma^{t}_{\phi \bar{\ell}}}{2} \right) \nonumber \\
    - &\frac{Y_{L}}{Y^{\text{eq}}_{L}} \left( \frac{\gamma_{D1}}{2}  + \gamma_{t_1}+\gamma_{t_2} + \gamma_{V_1} + \gamma_{V_3}  +  \gamma_{V_4}  +  \gamma_{V_5}+  \frac{\gamma_{\phi\psi}}{2} +  \frac{\gamma'_{\phi\psi}}{2} + \frac{\gamma^{t}_{\phi \ell}}{2} +  \frac{\gamma^{t}_{\phi \bar{\ell}}}{2}  + \gamma^{t}_{ \psi\bar \ell} \right. \nonumber \\ &\left. +\frac{Y_{N_1} }{Y_{N_1}^{\text{eq}}} (\gamma_s+\gamma_{V_2}+\gamma_{V_6})  
    \right),~{\rm and}  \label{eq:bz2} \\
H S z ~\frac{dY_{\Delta \psi}}{dz} &= \left( \frac{Y_{N}}{Y^{eq}_{N}} - 1\right) \epsilon_{DM} \gamma_{D_2} - {2\Theta^{{\phi \psi}}_{\epsilon_2}}   + {2\Theta^{{\phi^{\dagger} \bar \psi}}_{\epsilon'_2} }+  {2 \Theta^{{\phi \ell}}_{\epsilon^{t}_1}}   + {2 \Theta^{{\bar \ell \psi }}_{\epsilon^{t}_1}} +{2 \Theta^{{\phi \bar \ell  }}_{\epsilon^{t}_2}} +{2 \Theta^{{ \ell \psi }}_{\epsilon^{t}_2}} \nonumber \nonumber \\ & + \frac{Y_L}{Y^{eq}_L} \left(  \frac{\gamma_{\phi \psi}}{2}  - \frac{\gamma'_{\phi \psi}}{2} +\frac{\gamma^{t}_{\phi l}}{2}  - \gamma^t_{\bar \ell \phi}  \right) \nonumber \\ & - \frac{ Y_{\Delta \psi} }{Y^{eq}_{\psi}} \left(\frac{\gamma_{D_2} }{2}+\frac{\gamma_{\phi \psi}}{2} +\frac{\gamma'_{\phi \psi}}{2} + \frac{\gamma^{t}_{\phi l}}{2}  +\frac{\gamma^{t}_{\psi \bar l}}{2}    +\frac{\gamma^{t}_{\phi \bar l}}{2} + \frac{\gamma^{t}_{\psi l}}{2}  \right), 
 \label{eq:bz3}
\end{align}
where $\gamma_{D_1}$ and $\gamma_{D_2}$ are the thermal averaged decay widths of $N_1 \to \ell~H$ and $N_1 \to \phi~\psi$, respectively.
$\gamma_{t_1}$ is the thermal average of the processes  $N_1 ~u  \leftrightarrow d~\bar{\ell} $ and  $N_1 ~\bar{u}  \leftrightarrow \bar{d}~ \ell$, and  $\gamma_{t_2}$ is the same for the processes $N_1~ d  \leftrightarrow u ~\ell$ and $N_1 ~\bar{d}  \leftrightarrow \bar{u} ~\bar{\ell} $. We denote the sum of these contributions as $\gamma_t=\gamma_{t_1}+\gamma_{t_2}$. Similarly, $\gamma_{s, V}$, where $V=W_\mu, Z_\mu$, are the thermally averaged cross sections of scattering processes involving the annihilation of $N_1$. The quantities $\gamma_{\phi \psi}$, $\gamma'_{\phi \psi}$, $\gamma^{t}_{\phi \ell}$, $\gamma^{t}_{\phi \bar\ell}$, $\gamma^{t}_{\psi \bar\ell}$, $\gamma_{LL}$, $\gamma_{HL}$ and $\gamma^{t}_{\psi \ell}$ represent the thermally averaged scattering cross-sections of the processes $\phi \psi \to \ell H$, $\phi^{\dagger} \bar \psi \to \ell H$, $\phi \ell \to \psi H$, $\phi \bar \ell \to \psi \bar H$, $\psi \bar \ell \to \phi \bar H$, $\ell \ell \to \bar H \bar H$, $H \ell \to \bar H \bar \ell$  and  $\psi \ell \to \phi H$, respectively (as given in the Appendex~\ref{eq:apendex:thermal:average}). While $\Theta^{{a b}}_{\epsilon_{\alpha}}$, given in Appendix~\ref{eq:apendex:thermal:averageep}, represent the thermal average of $\epsilon_{\alpha}\sigma^{ab}$ corresponding to  the respective scattering process. In Eq. \ref{eq:bz2}, the terms proportional to $Y_{\Delta \psi} $ indicate the transfer of dark-sector asymmetry into the lepton asymmetry. 
Similarly, in Eq.~\ref{eq:bz3}, the terms involving $Y_{L}$ represent the transfer of visible sector asymmetry into the dark sector. The strength of the transfer of asymmetry from the dark sector to the visible sector and vice versa depends on the choice of parameter values. In a simplified scenario, if the scattering processes involving dark matter occur in such a way that the numbers of dark sector particles $\psi$ and $\bar{\psi}$ remain equal during the evolution, then no dark sector asymmetry is generated, nor can any dark-sector asymmetry be transferred to the visible sector. Under this condition, only two equations are required to analyze the resulting lepton asymmetry. 
As described below, this case can be solved analytically in a schematic manner.
\subsubsection{Case when asymmetry in dark sector is absent} 
We shall explore the specific case when dark sector asymmetry is negligible. Consequently, the abundances of $\psi$ and $\bar{\psi}$ remain equal ($Y_{\psi}(z) = Y_{\bar{\psi}}(z)$).
Under this condition, Eq.~\ref{eq:bz1} and Eq. \ref{eq:bz2} simplify to
\begin{align}
    \frac{dY_{N_1}}{dz} + P_{N}(z) Y_{N_1} &= Q_{N}(z), \label{eq:bzf1} \\
    \frac{dY_{L}}{dz} + P_{L}(z) Y_{L} &= Q_{L}(z), \label{eq:bzf2}
\end{align}
where  $P_N(z)$, $Q_N(z)$, $P_L(z)$ and $Q_L(z)$ are defined in Eq. \ref{eq:pz}, \ref{eq:Qz}, \ref{eq:Pz} and \ref{eq:Qz_decomp1} respectively. 
 \begin{align}
P_{N}(z) &= \frac{\gamma_{D_1} + \gamma_{D_2} + 2\gamma_s + 2\gamma_{t_1} + 2\gamma_{t_2} + 2\gamma_V}{H S z \, Y_{N_1}^{\rm eq}}, \label{eq:pz} \\[10pt]
Q_{N}(z) &= Y_{N_1}^{\rm eq} \, P_{N}(z), \label{eq:Qz} \\[10pt]
P_{L}(z) &= \frac{1}{H S z \, Y_L^{\rm eq}} \Bigg[
\frac{\gamma_{D_1}}{2} + \gamma_{t_1} + \gamma_{t_2} + \gamma_{V_1} + \gamma_{V_3} + \gamma_{V_4} + \gamma_{V_5} \nonumber \\
&+ \frac{\gamma_{\phi\psi}}{2} + \frac{\gamma'_{\phi\psi}}{2} 
+ \frac{\gamma^t_{\phi \ell}}{2} + \frac{\gamma^t_{\phi \bar{\ell}}}{2} 
+ \gamma^t_{\bar{\ell} \psi} 
+ \frac{Y_{N_1}}{Y_{N_1}^{\rm eq}} (\gamma_s + \gamma_{V_2} + \gamma_{V_6})
\Bigg] , ~{\rm and}  \label{eq:Pz} \\[10pt]
Q_{L}(z) &= \frac{\epsilon_1 \gamma_{D_1}}{H S z} \left( \frac{Y_{N_1}}{Y_{N_1}^{\rm eq}} - 1 \right)
+ \frac{2}{H S z} \Big[{\Theta^{{\phi \psi}}_{\epsilon_2}}   + {\Theta^{{\phi^{\dagger} \bar \psi}}_{\epsilon'_2} } - { \Theta^{{\phi \ell}}_{\epsilon^{t}_1}} - { \Theta^{{\bar \ell \psi }}_{\epsilon^{t}_1}}  +{ \Theta^{{\phi \bar \ell  }}_{\epsilon^{t}_2}} + { \Theta^{{ \ell \psi }}_{\epsilon^{t}_2}}\Big] \nonumber \\[8pt]
&= \epsilon_1 Q_1(z) + Q_{\epsilon_2}(z) +  Q_{\epsilon'_2}(z) 
+  Q_{\epsilon^t_1}(z) +  Q'_{\epsilon^t_1}(z) 
+  Q_{\epsilon^t_2}(z) +  Q'_{\epsilon^t_2}(z) . \label{eq:Qz_decomp1}
\end{align}
In Eq. \ref{eq:Qz_decomp1}, the individual source terms ${Q}_i(z)$ are defined implicitly through decomposition. This simple form, allows us to solve Eq. \ref{eq:bzf1} and Eq. \ref{eq:bzf2} formally with the solutions  
\begin{align}
    Y_{N_1}(z) &= \frac{1}{\mu(z)} \left[ Y^{\rm eq}_{N_1}(z_0) + \int_{z_0}^z \mu(s) \, Q_{N}(s) \, ds \right],~{\rm and} \label{eq:yn1} \\
    Y_L(z) &= \epsilon_1 \kappa_1 + \kappa_2 + \kappa_3 + \kappa_4 + \kappa_5 + \kappa_{6} + \kappa_{7}, \label{eq:yl1}
\end{align}
respectively, where
\(
    \mu(z) = \exp\left( \int_{z_0}^z P_{N}(z') \, dz' \right).
    \)
The definitions of the various efficiency factors $\kappa_i$ are given by
\begin{align}
    \kappa_1 &= \frac{1}{\nu(z)} \int_{z_0}^z \nu(z') \, Q_1(z') \, dz', &
    \kappa_2 &= \frac{1}{\nu(z)} \int_{z_0}^z \nu(z') \, Q_{\epsilon_2}(z') \, dz', \nonumber\\
    \kappa_3 &= \frac{1}{\nu(z)} \int_{z_0}^z \nu(z') \, Q_{\epsilon'_2}(z') \, dz', &
    \kappa_4 &= \frac{1}{\nu(z)} \int_{z_0}^z \nu(z') \, Q_{\epsilon^{t}_{1}}(z') \, dz', \nonumber\\
    \kappa_5 &= \frac{1}{\nu(z)} \int_{z_0}^z \nu(z') \, Q'_{\epsilon^{t}_{1}}(z') \, dz', &
    \kappa_6 &= \frac{1}{\nu(z)} \int_{z_0}^z \nu(z') \, Q_{\epsilon^{t}_{2}}(z') \, dz', \nonumber\\
    \kappa_7 &= \frac{1}{\nu(z)} \int_{z_0}^z \nu(z') \, Q'_{\epsilon^{t}_{2}}(z') \, dz',
\end{align}
where each $\kappa_i$ represents the accumulated contribution of the corresponding source term ${Q}_i(z)$, while being weighted by the integrating factor $\nu(z)$ given by
\begin{align}
    \nu(z) &= \exp\left( \int_{z_0}^z P_{L}(z') \, dz' \right).
\end{align}
{These efficiency factors encode the different sources of matter-antimatter asymmetry. Specifically, $\kappa_1$ contributes to the asymmetry only when $Y_{N_1}$ deviates from its equilibrium value $Y^{\rm eq}_{N_1}$. The factors $\kappa_2$ and $\kappa_3$ arise from the $s$-channel $2\to 2$ scattering processes $\phi \psi \to \ell H$ and $\phi^\dagger \bar{\psi} \to \ell H$, respectively; their contributions are maximized when the $C\!P$-violating combinations $\epsilon_2 \sigma^{\phi \psi}$ and $\epsilon'_2 \sigma^{\phi^\dagger \bar{\psi}}$ are largest. Similarly, $\kappa_4$ and $\kappa_5$ originate from the $t$-channel processes $\ell \phi \to H \psi$ and $\bar{\ell} \psi \to \phi H^\dagger$, respectively, while $\kappa_6$ and $\kappa_7$ stem from the related $t$-channel processes $\bar{\ell} \phi \to H^\dagger \psi$ and $\ell \psi \to H \phi$, respectively. The contributions from  $\kappa_4$, $\kappa_5$, $\kappa_6$, and $\kappa_7$ are maximized when the respective $C\!P$-violating scattering rates $\epsilon^t_1 \sigma^{\phi \ell}$, $\epsilon^t_1 \sigma^{\bar{\ell} \psi}$, $\epsilon^t_2 \sigma^{\phi \bar{\ell}}$, and $\epsilon^t_2 \sigma^{\ell \psi}$ are the largest. Finally, the $C\!P$ asymmetry parameter $\epsilon_1$ and all efficiency factors $\kappa_{1,2,3,4,5,6,7}$ depend on the Yukawa couplings $y_{1,2,3}$, the heavy neutrino masses $m_{N_{1,2,3}}$, and the DM masses $m_\phi$ and $m_\psi$. Consequently, the lepton asymmetry $Y_L$ depends solely on these free parameters.} 

In Eq.~\ref{eq:bz2} and Eq. \ref{eq:bz3}, the parameters $\epsilon_2$, $\epsilon'_2$, $\epsilon^t_1$, and $\epsilon^t_2$ appear linearly. The $C\!P$ asymmetry parameters $\epsilon_{1}$ and $\epsilon_{DM}$ are associated with the decays of RHNs, which behave as a source term for the visible sector and dark sector asymmetry, respectively. While the scattering processes serves as a source to generate asymmetry in both the dark sector and the visible sector, it also acts as a washout process, as well as transferring asymmetry from the dark sector to the visible sector and vice-versa. The $C\!P$ asymmetry sources $\epsilon_2$ and $\epsilon'_2$ contribute inversely to each other in the leptonic sector. 
\section{Numerical Analysis and results}
\label{sec:numericalanalysis}
In this section, we perform a numerical analysis of the lepton asymmetry (visible sector) and DM asymmetry generated by decays and scattering processes involving the heavy right-handed neutrinos. These asymmetries depend on the free parameters $y_i$, $m_{N_i}$ (for $i=1,2,3$), $m_\phi$, and $m_\psi$. The $C\!P$ asymmetry parameters discussed in Sec.~\ref{sec:CPasymmetry} play a crucial role in the generation of these asymmetries. In this framework, there are eight such parameters and we have considered only $\epsilon_1$, $\epsilon_{DM}$, $\epsilon_2$, $\epsilon'_2$, $\epsilon^t_1$, and $\epsilon^t_2$ in our investigation as the $C\!P$ asymmetries from three-body and four-body decays are not allowed here due to mass degeneracy in the dark sector ($m_\phi=m_\psi$).  The $C\!P$ asymmetry parameter $\epsilon_1$ and $\epsilon_{DM}$ depend on $y_i$ and $m_{N_i}$, while other $C\!P$ asymmetry parameters $\epsilon_2$, $\epsilon'_2$, $\epsilon^{t}_1$ and $\epsilon^{t}_2$ depend on the decay width of $N_1$ in addition to $y_i$, and $m_{N_i}$. We discuss the visible sector asymmetry study in sub-section~\ref{sec:numerical:visible} and dark sector asymmetry study in sub-section~\ref{sec:numerical:dark} below. 
\subsection{Numerical Analysis of visible sector asymmetry} 
\label{sec:numerical:visible}
\begin{figure}[!htp]
    \centering
\begin{subfigure}{0.32\textwidth}
        \centering
        \caption*{(a)}
\includegraphics[width=\textwidth,height=\textwidth]{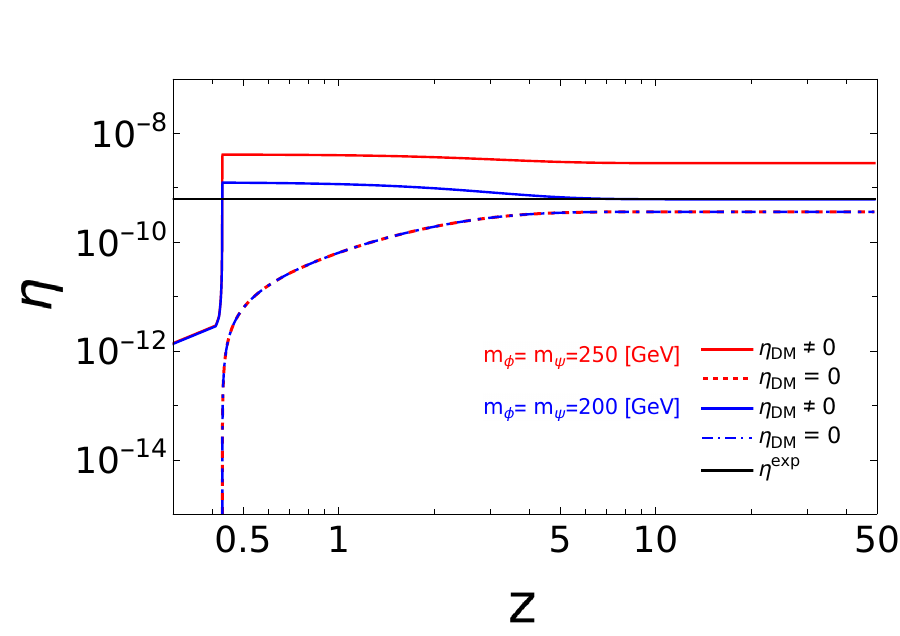 }
    \end{subfigure}
    \hfill
    \begin{subfigure}{0.32\textwidth}
        \centering
        \caption*{(b)}
\includegraphics[width=\textwidth,height=\textwidth]{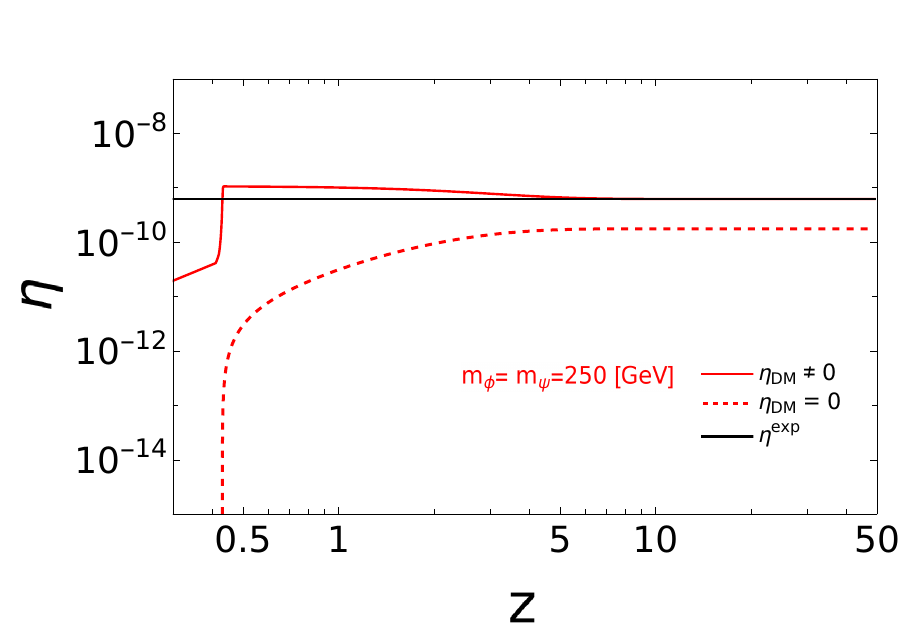 }
    \end{subfigure}
    \hfill
    \begin{subfigure}{0.32\textwidth}
        \centering
        \caption*{(c)}
\includegraphics[width=\textwidth,height=\textwidth]{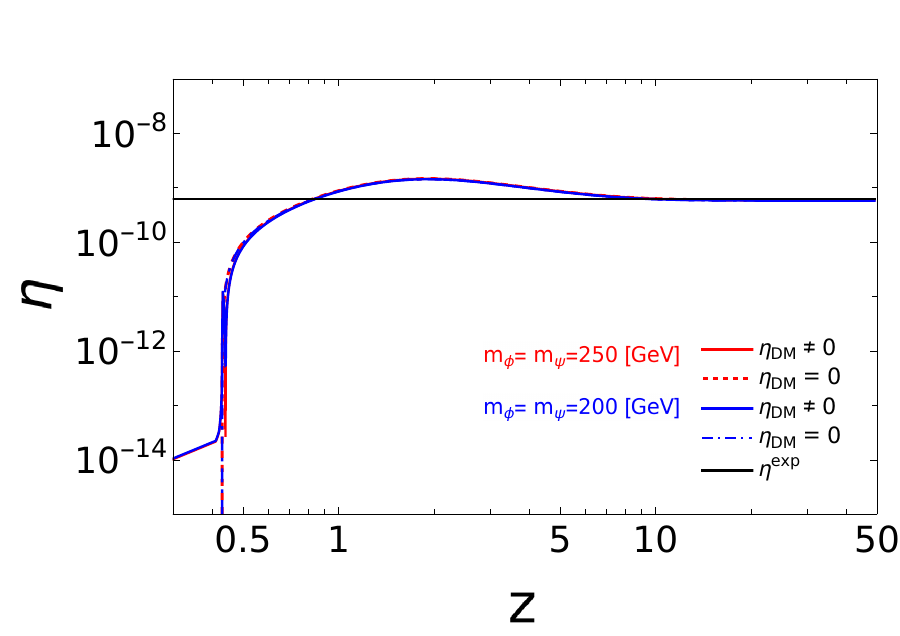 }
    \end{subfigure}
\caption{The baryon asymmetry parameter $\eta$ as a function of the dimensionless variable $z = \frac{m_{N_1}}{T}$. The parameter values considered for these studies are given in Table~\ref{tab:3benchmark-paramseta}}
\label{fig:etavsz}
\end{figure}

\begin{figure}[!htp]
    \centering

    \begin{subfigure}{0.49\textwidth}
        \centering
        \caption*{(a)}
\includegraphics[width=\textwidth,height=\textwidth]{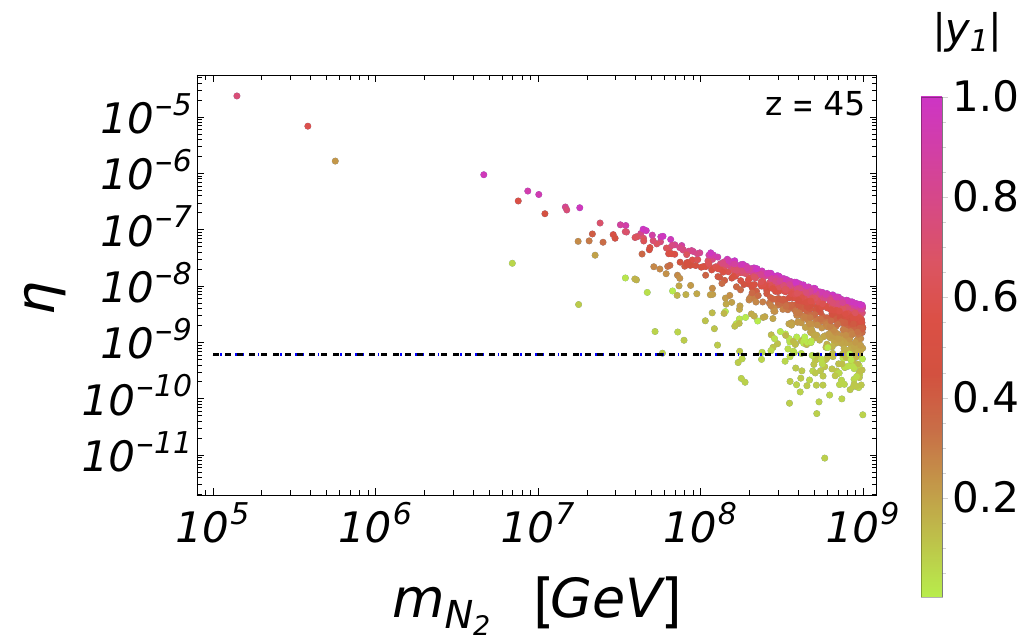}
    \end{subfigure}
    \hfill
    \begin{subfigure}{0.49\textwidth}
        \centering
        \caption*{(b)}
\includegraphics[width=\textwidth,height=\textwidth]{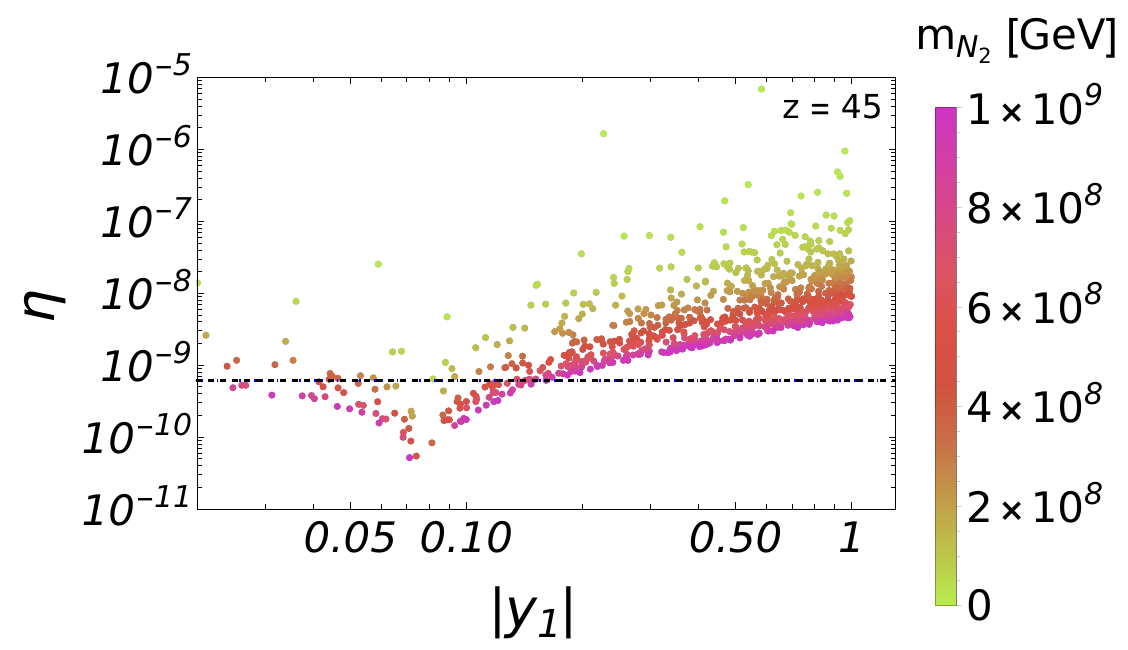}
    \end{subfigure}

    \caption{
    Scan of the baryon asymmetry parameter $\eta[z=45]$, 
    (a) as a function of the right-handed neutrino mass $m_{N_2}$, with color variation representing the Yukawa coupling $|y_1|$, 
    (b) as a function of $|y_1|$, with color variation representing $m_{N_2}$. The parameter values considered for the scan are given in Table~\ref{tab:benchmark-paramseta}.
    }
    \label{fig:Scan:etavsmn2y11a}
\end{figure}

As explained in the previous sections, in the present framework, the lepton asymmetry is influenced by the dark sector in multiple ways. Our numerical study elaborated in this section will categorically establish that this influence is significant enough to have sufficiently large leptogenesis with TeV-scale $N_1$. We shall illustrate this with the results presented in Fig. \ref{fig:etavsz}, were the baryon number asymmetry, $\eta$ is given as a function of $z=\tfrac{m_{N_1}}{T}$. Parameter values used for this study are given in Table~\ref{tab:3benchmark-paramseta}. The couplings, $y_2$ and $y_3$ as well as the masses of the heavy neutrinos, $N_2$ and $N_3$ are fixed in both the cases described in the two plots.
In Fig.~\ref{fig:etavsz}(a), the observed baryon asymmetry is achievable for $m_\phi=m_\psi=200$ GeV for a corresponding $y_1=0.0059 (1-i)$. However, notice that this case correspond to  over-abundant dark matter scenario. The red curve corresponding to $m_\phi=m_\psi=250$  GeV for the same coupling produces larger than observed baryon asymmetry, while being consistent with the dark matter abundance. Fig.~\ref{fig:etavsz}(a) demonstrates the significance of dark matter asymmetry ($\eta_{DM}$) in leptogenesis. As expected, the when $\eta_{DM}=0$, the baryon asymmetry is insensitive to the dark matter mass. One can conclude that, for $m_{N_1}=2$ TeV, and for the other right-handed neutrino masses and seesaw Yukawa's fixed as in this case, the dark matter mass around or below 200 GeV are not consistent with both dark matter observations and required baryon asymmetry. Keeping the dark matter mass a little higher, at 250 GeV, it is possible to adjust the coupling $y_1$ to get the right value of $\eta$, while being consistent with the dark matter observations, as illustrated in Fig.~\ref{fig:etavsz}(b). The coupling required in this case is $y_1=\tfrac{0.1}{\sqrt{2}}(1-i)$.  
Fig.~\ref{fig:etavsz}(c) demonstrates that the dark matter asymmetry is quite sensitive to the value of $m_{N_2}$. At $m_{N_2}={\cal O}(10^7)$ GeV, $\eta_{DM}$ is negligible. However, suitable couplings provide the required baryon asymmetry, while being consistent with the dark matter observations at for dark matter mass of 250 GeV. 
\begin{table}[!htb]
    \centering
    \begin{tabular}{c|c|c|c|c|c|c }
        \hline\hline
         B.P. &$m_{N_1}$ [GeV] & $m_{N_2}$ [GeV] & $m_{N_3}$ [GeV] & $y_1$ & $y_{2}$ & $y_{3}$ \\
        \hline
     (a)  & $2000$ &$10^9$ &$26 \times 10^{12}$ &$0.0059(1-i)$ & $0.20(1+i)$ & $0.3 \sqrt{2\pi}(1-i)$ \\
        \hline
    (b)  & $2000$ &$10^9$ &$26 \times 10^{12}$ &$\frac{0.1}{\sqrt{2}}(1-i)$ & $0.20(1+i)$ & $0.3 \sqrt{2\pi}(1-i)$ \\
        \hline  
        (c)  & $2000$ &$1.5 \times10^7 $ &$26 \times 10^{12}$ &$0.12(1-i)$ & $0.20(1+i)$ & $0.3 \sqrt{2\pi}(1-i)$ \\
        \hline\hline 
    \end{tabular}
\caption{Fixed parameter points for $\eta$ vs z plot Fig~\ref{fig:etavsz}. }
    \label{tab:3benchmark-paramseta}
\end{table}
Although leptogenesis depends on the parameters $y_1,y_2,y_3,m_{\phi},m_{\psi},m_{N_1},m_{N_2},m_{N_3}$, the resulting asymmetry $\eta$ is particularly sensitive to just two of them: $y_1$ and $m_{N_2}$. Consequently, a broad range of $(y_1,m_{N_2})$ values can reproduce the observed $\eta$. This is explicitly shown in the scans over the parameter space as  presented in Fig.~\ref{fig:Scan:etavsmn2y11a},
Fig. \ref{fig:scan:DMAsymmetryt}, and Fig. \ref{fig:Scan:etavsmn2y11b}.

\begin{table}[!htb]
    \centering
    \begin{tabular}{c|c|c|c|c|c}
        \hline\hline
        $m_{N_1}$ [GeV] & $m_{N_3}$ [GeV]& $m_{\phi}$[GeV] & $m_{\psi}$[GeV] & $y_{2}$ & $y_{3}$ \\
        \hline
        $2000$ & $26 \times 10^{12}$ & 250 & 250 & $0.200781(1+i)$ & $0.3 \sqrt{2\pi}(1-i)$ \\
        \hline\hline
        \multicolumn{3}{c|}{$8 \times 10^4 < m_{N_2} < 10^9 $}&\multicolumn{3}{c}{$\tfrac{10^{-3}}{\sqrt{2}}(1-i) < y_1 < \tfrac{1}{\sqrt{2}}(1-i)$}\\ \hline\hline
    \end{tabular}
\caption{Parameter values considered for the scan results presented in Fig.~\ref{fig:Scan:etavsmn2y11a}, Fig.~\ref{fig:scan:DMAsymmetryt}, and Fig.~\ref{fig:Scan:etavsmn2y11b}. ~~~~~~~~~~~~~~~~~~~~~~~~~~~~~~~~~~~~~~~~~~~~~~~~~~~~~~~~~~~~~~~~~~~~~~~~~~~~~~~~~~~~~~~}
    \label{tab:benchmark-paramseta}
\end{table}
 The scans were carried out over the ranges \(  \frac{10^{-3}(1-i)}{\sqrt{2}} < y_1 < \frac{(1-i)}{\sqrt{2}} \) and \( 8 \times 10^4~\mathrm{GeV} < m_{N_2} < 10^9~\mathrm{GeV} \), while other parameter were fixed at values given in Table~\ref{tab:benchmark-paramseta}. For each point in this parameter space, we solved the Boltzmann Eq.s~(\ref{eq:bz1})--(\ref{eq:bz3}) and extracted the corresponding value of \( \eta \) at \( z = 45 \).  The left panel in Fig.~\ref{fig:Scan:etavsmn2y11a} shows $\eta$ as a function of $m_{N_2}$ with the color scale encoding $|y_1|$. The right panel  in Fig.~\ref{fig:Scan:etavsmn2y11a} shows $\eta$ as a function of $|y_1|$, with the color scale encoding $m_{N_2}$. Both panels include horizontal lines marking the observed value of the BAU $\eta_{\rm obs}$ in $3~\sigma$ limit. Corresponding to a large value of $m_{N_2}$, a combination with a small value of $y_1$ can provide the observed value of $\eta$. It is evident that $\eta \propto \frac{1}{m_{N_2}}$ with $\eta > 0$. This behavior typically arises from the $C\!P$ asymmetry parameters. For a large value of $m_{N_2}$, a smaller value of $y_1$ leads to a smaller value of $\eta$, whereas a larger value of $y_1$ results in a larger value of $\eta$. For small values of $|y_1|$, $\eta$ remains suppressed, whereas it increases significantly at larger values of $y_1$.
\subsection{Numerical Analysis of Dark Sector Asymmetry} 
\label{sec:numerical:dark}
\begin{figure}[!htp]
    \centering
    \begin{subfigure}{0.31\textwidth}
        \centering
        \caption*{(a)}
\includegraphics[width=\textwidth,height=\textwidth]{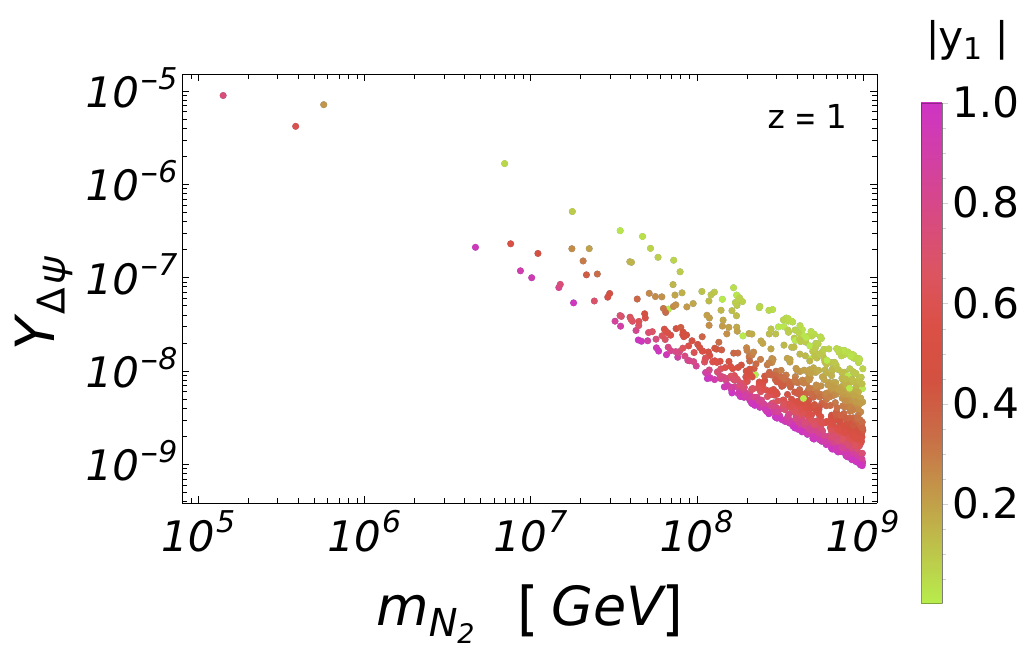}
    \end{subfigure}\hfill
    \begin{subfigure}{0.31\textwidth}
        \centering
        \caption*{(b)}
\includegraphics[width=\textwidth,height=\textwidth]{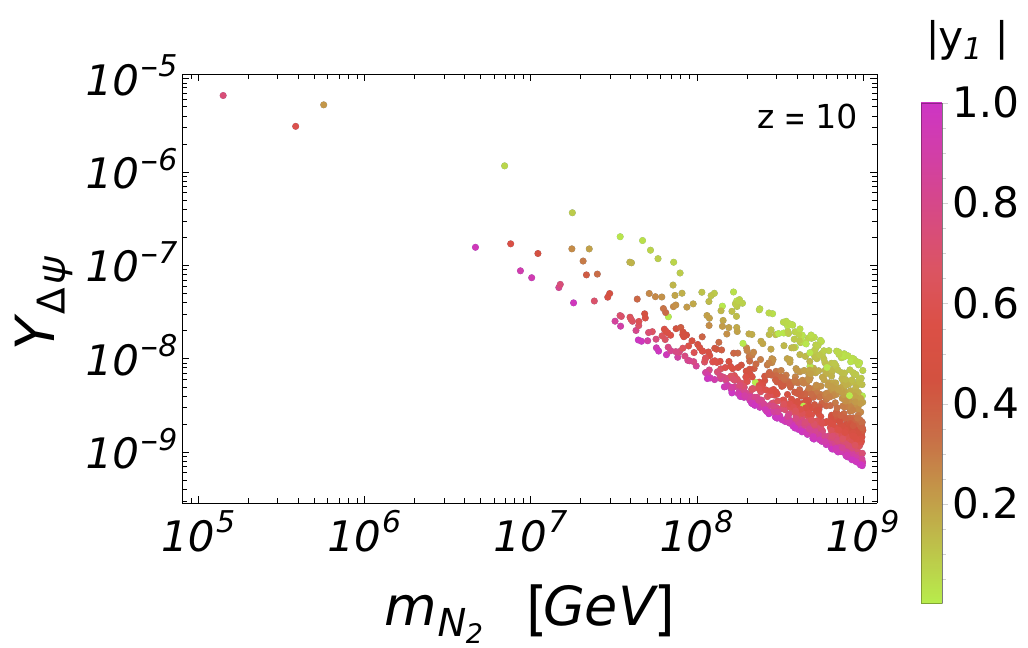}
    \end{subfigure}\hfill
    \begin{subfigure}{0.31\textwidth}
        \centering
        \caption*{(c)}
\includegraphics[width=\textwidth,height=\textwidth]{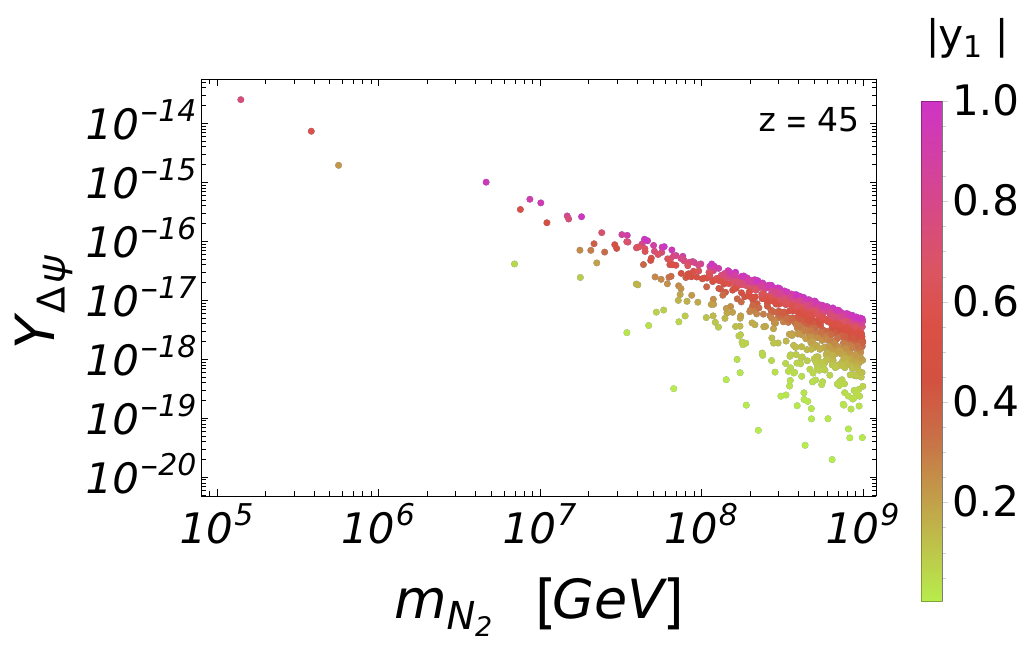}
    \end{subfigure}

    \vspace{0.5em}

    \begin{subfigure}{0.31\textwidth}
        \centering
        \caption*{(d)}
\includegraphics[width=\textwidth,height=\textwidth]{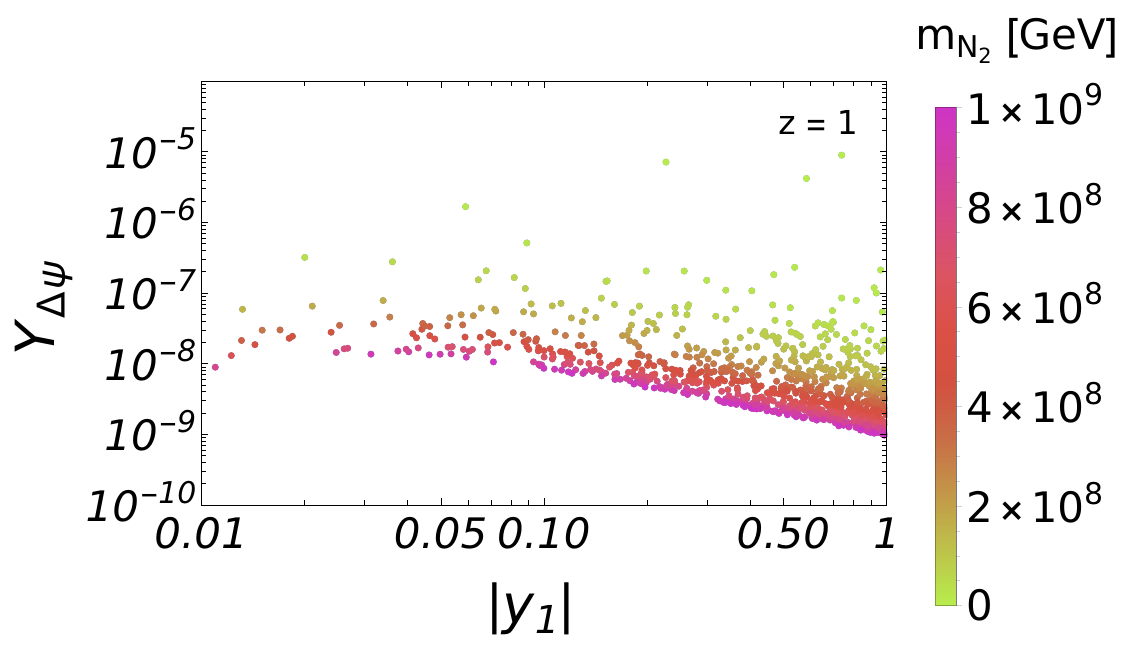}
    \end{subfigure}\hfill
    \begin{subfigure}{0.31\textwidth}
        \centering
        \caption*{(e)}
\includegraphics[width=\textwidth,height=\textwidth]{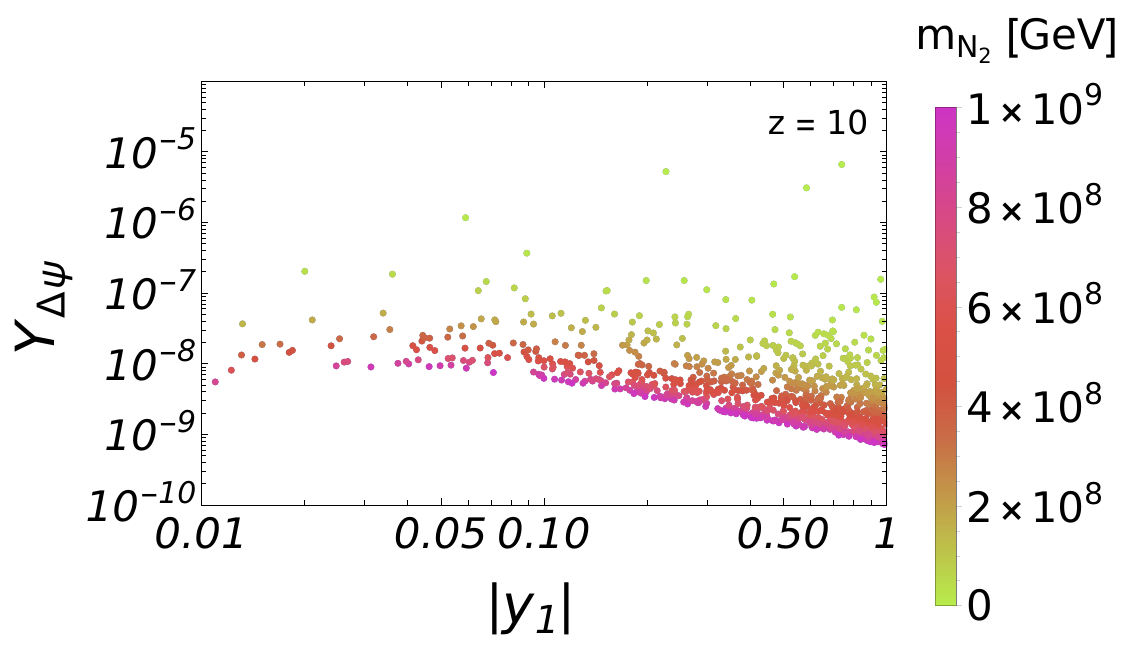}
    \end{subfigure}\hfill
    \begin{subfigure}{0.31\textwidth}
        \centering
        \caption*{(f)}
\includegraphics[width=\textwidth,height=\textwidth]{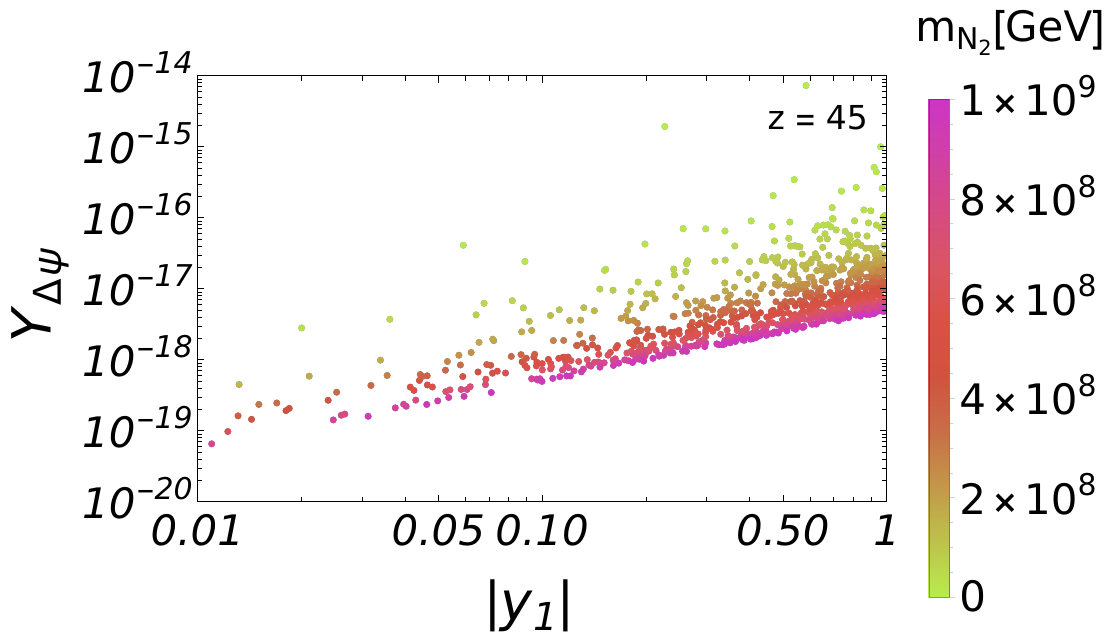}
    \end{subfigure}

    \caption{Dark-sector asymmetry \( Y_{\Delta \psi} \): 
    (a)–(f) as described in the text, for the benchmark point given in Table~\ref{tab:benchmark-paramseta}.}
    \label{fig:scan:DMAsymmetryt}
\end{figure}

We investigate the numerical behavior of the dark sector asymmetry and its possible impact on the visible-sector asymmetry. Our analysis indicates that, for certain choices of parameters, the dark sector asymmetry can significantly affect the visible sector asymmetry. This is particularly so when the dark sector asymmetry is sizable at early times (small values of \( z \)). For other regions of parameter space, however, the visible sector asymmetry remains largely insensitive to the dark-sector contribution. To study this interplay, we compute the dark sector asymmetry \( Y_{\Delta \psi} \) at three representative values of the evolution parameter \( z \): \( z = 1 \), \( z = 10 \), and \( z = 45 \), as shown in Fig.~\ref{fig:scan:DMAsymmetryt}. The parameter values used in this analysis are listed in Table~\ref{tab:benchmark-paramseta}. For each point in parameter space, we numerically solve Eq.~\ref{eq:bz3} to obtain the corresponding values of \( Y_{\Delta \psi} \) at the three chosen values of \( z \). We then perform parameter scans of \( Y_{\Delta \psi} \) as a function of \( m_{N_2} \) and \( |y_1| \) at \( z = 1 \), \( z = 10 \), and \( z = 45 \), as illustrated in Fig.~\ref{fig:scan:DMAsymmetryt}.

The dark-sector asymmetry at \( z = 1 \) and \( z = 10 \) is both sizable and nearly identical. In contrast, at later times (\( z = 45 \)) the asymmetry becomes strongly suppressed. For all values of \( z \), the asymmetry \( Y_{\Delta \psi} \) increases as the mass of the right-handed neutrino \( m_{N_2} \) decreases. The dependence of \( Y_{\Delta \psi} \) on the Yukawa coupling \( |y_1| \), however, varies with \( z \) and exhibits distinct behaviors at different stages of the evolution.
In this analysis, the mass of the lightest right-handed neutrino is fixed at \( m_{N_1} = 2000~\mathrm{GeV} \). When the mass \( m_{N_2} \) approaches \( m_{N_1} \), the \( C\!P \)-asymmetry parameter \( \epsilon_{\text{DM}} \) increases sharply, leading to an enhancement of the dark sector asymmetry. As \( m_{N_2} \) moves further away from \( m_{N_1} \), the dark sector asymmetry decreases steadily, as shown in subfigures (a), (b), and (e) of Fig.~\ref{fig:scan:DMAsymmetryt}.
An increase in the Yukawa coupling \( y_1 \) enhances both the decay rate of the process \( N_1 \to \phi\,\psi \) and the associated scattering cross sections. As a result, the dark sector asymmetry initially grows because the parameter \( \epsilon_{\text{DM}} \) itself increases with \( y_1 \). Beyond a critical value of \( y_1 \), however, washout processes become dominant and suppress the asymmetry. This leads to a characteristic turnover, after which the dark sector asymmetry decreases, as seen in subfigures (d), and (e) of Fig.~\ref{fig:scan:DMAsymmetryt}.
\begin{figure}[!htp]
    \centering
 \begin{subfigure}{0.31\textwidth}
        \centering
        \caption*{(a)}
        \includegraphics[width=\textwidth,height=\textwidth]{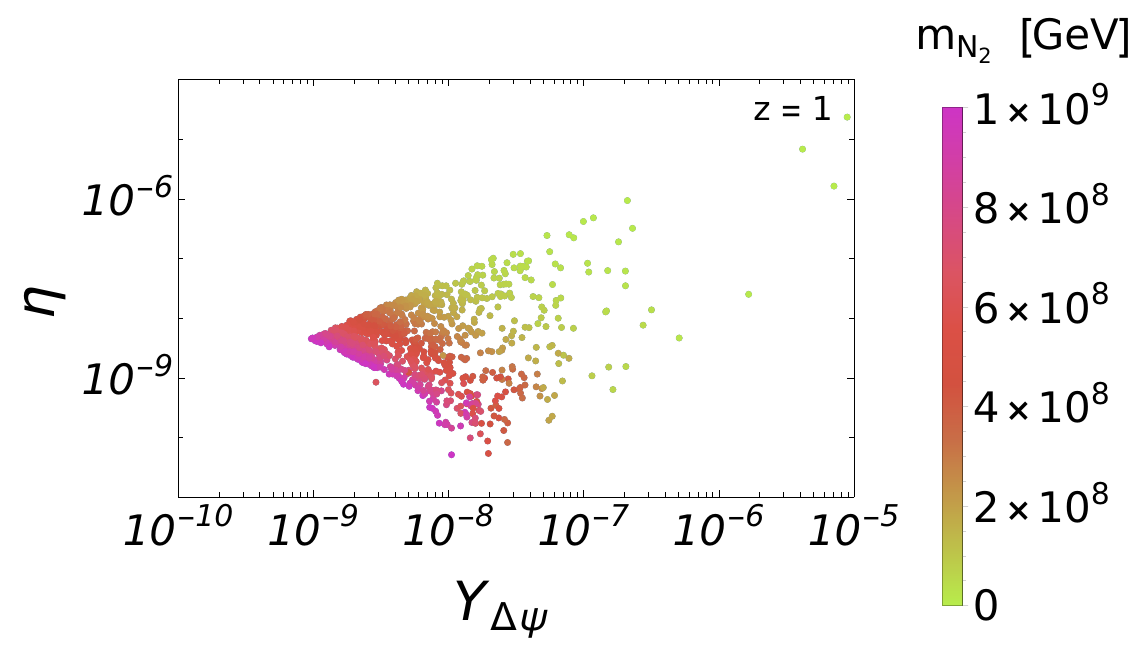}
    \end{subfigure}
    \hfill
    \begin{subfigure}{0.31\textwidth}
        \centering
        \caption*{(b)}
        \includegraphics[width=\textwidth,height=\textwidth]{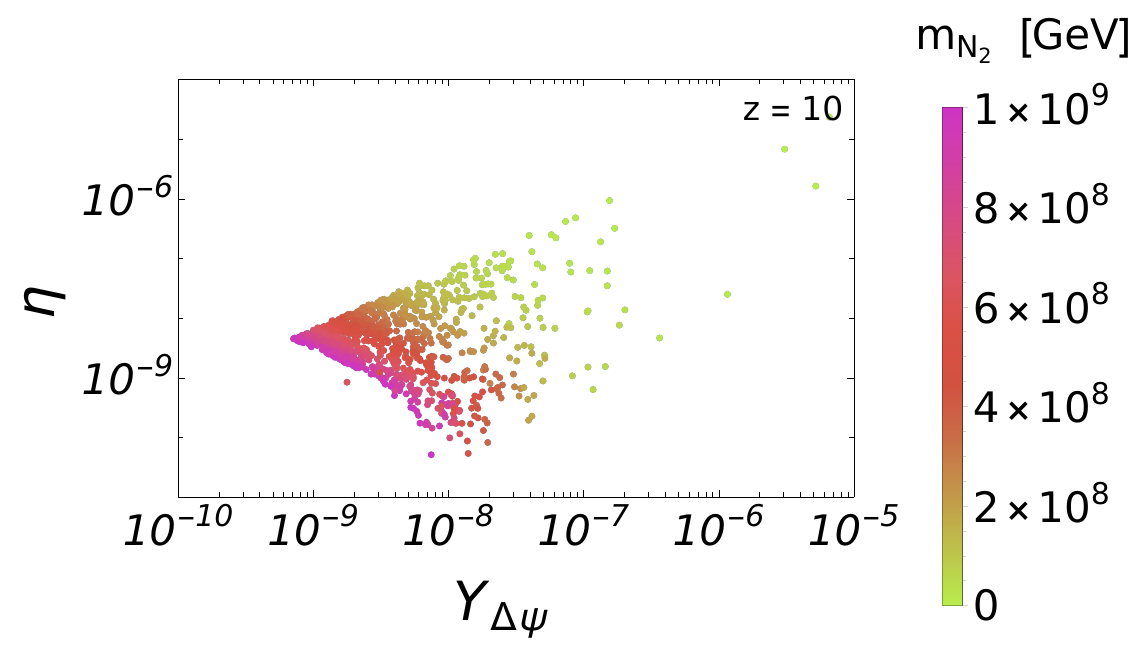}
    \end{subfigure}
    \hfill
        \begin{subfigure}{0.31\textwidth}
        \centering
        \caption*{(c)}
        \includegraphics[width=\textwidth,height=\textwidth]{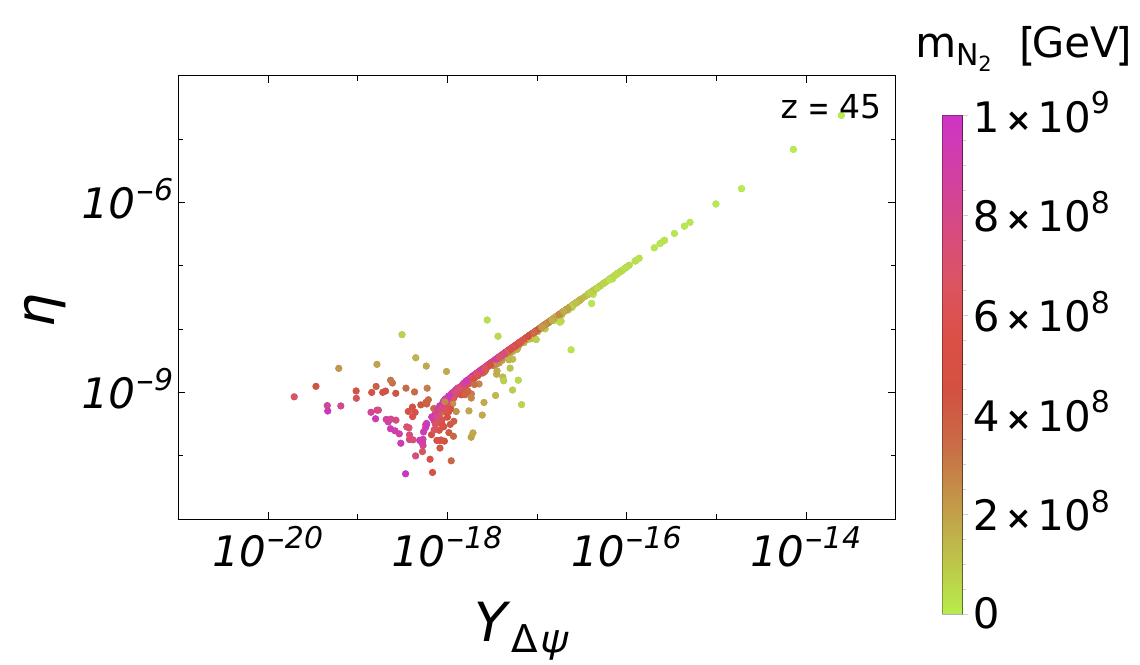}
    \end{subfigure}
\caption{Scan of the dark sector asymmetry \( Y_{\Delta \psi} \) as a function of \( \eta \) for an alternative benchmark point is given in Table~\ref{tab:benchmark-paramseta}.
    }
    \label{fig:Scan:etavsmn2y11b}
\end{figure}

Although the dark sector asymmetry can affect the visible sector asymmetry, as illustrated in Fig.~\ref{fig:etavsz} and further supported by the scatter plot in Fig.~\ref{fig:Scan:etavsmn2y11b}, the transfer of asymmetry from the dark sector to the visible sector is most efficient at early times, namely at \( z = 1 \) and \( z = 10 \). At later times (\( z = 45 \)), the transfer becomes highly suppressed and has a negligible impact on the dark matter relic density and visible sector asymmetry. This is because dark matter decouples at a later epoch, by which time the number densities of \( \psi \) and \( \bar{\psi} \) are approximately equal.
\section{\label{sec:relultandconclusion}Conclusion}

In this work, we present an analysis of a simple extension of the SM that accommodates both a DM candidate and a viable mechanism for leptogenesis in the early Universe. The model introduces three Majorana right-handed neutrinos, a $Z_2$-odd real scalar field $\phi$, and a charged neutral fermion $\psi$. In this setup, both $\phi$ and $\psi$ are stable and serve as stable dark matter particles. To distinguish the scotino ($\psi$) from its antipartile, we introduce a global $U(1)_{L'}$ corresponding to a generalised lepton number.
The fermionic dark matter candidate $\psi$ interacts with other fields only through its Yukawa couplings. In contrast, the scalar field $\phi$ couples through quartic scalar couplings as well. However, the quartic coupling involved must be extremely small in order to satisfy dark matter direct detection constraints. As a consequence, $\phi$ cannot serve as a viable single-component dark matter candidate in this scenario, when considered as a WIMP. Since $\psi$ pair annihilation is prohibited when $\phi$ and $N_1$ are heavier, the only phenomenologically viable dark matter configuration in this model is a two-component fermion–scalar dark matter scenario, where the correct relic abundance is achieved predominantly through co-annihilation channels. The Yukawa coupling of $\phi$, $\psi$ interaction and the masses of dark matter candidate, masses of right handed neutrinos is constrained from the observed relic density. Minimum masses of dark matter candidate depends on the masses of right handed neutrinos. For simplicity, we have considered an exact mass degeneracy of $\psi$ and $\phi$ in our study.
For a lightest right-handed neutrino mass of $2000~\text{GeV}$, the lightest dark matter mass can naturally lie around $200~\text{GeV}$. Therefore, in our analysis we do not consider dark matter masses below $200~\text{GeV}$.

Canonical thermal leptogenesis with hierarchical right-handed neutrinos typically requires a very heavy right handed neutrino in order to generate the observed baryon asymmetry of the Universe, unless at least two of them are nearly degenerate in mass. In our model, however, additional possibilities arise due to the Yukawa interactions between the right-handed neutrinos and the dark matter candidate, which contribute to the leptogenesis mechanism.
In addition, the decay of $N_1$ into the dark matter candidate produces a primordial dark sector asymmetry. We have shown that this asymmetry can be efficiently transferred to the visible sector, thereby contributing to the observed baryon asymmetry. In the early universe, the dark sector asymmetry is significantly larger and capable of sourcing the visible sector asymmetry through the transfer mechanism. However, in the present epoch, the residual dark sector asymmetry becomes negligible, ensuring that the dark matter relic density remains unaffected.

\section{Acknowledgements}
\raggedright
PP and Avnish thank the DST, India for financial support with SERB CRG (CRG/2022/002670).
\par
\appendix
\section{{\label{decayandscatteringprocess}}The relevant Decay and scattering processes}
 Decay, inverse decay, and scattering processes are essential keys to generating DM and lepton asymmetry. Mostly DM asymmetry and lepton asymmetry are tracking with the evolution of \(N_1\), \(\phi\), \( \psi\), and \( \bar \psi\). The decay processes are sources which only generate the asymmetry, while scattering processes work as sources of asymmetry, washouts, and asymmetry transfer terms. The relevant processes to the Boltzmann equations \ref{eq:bz1}, \ref{eq:bz2}, and \ref{eq:bz3} of \(N_1\), \(\phi\), \( \Delta \psi = (\psi- \bar{\psi}) \), are listed as follows:

\begin{enumerate}
    \item Decay and inverse decay:\\ 
    $N_1 \leftrightarrow \ell~ H$, \quad  $N_1 \leftrightarrow
 \bar{\ell} ~H^{\dagger}$,    \quad  $N_1 \leftrightarrow \phi ~\psi$, \quad $N_1 \leftrightarrow
 \bar \phi ~\bar \psi$ 
  \end{enumerate}
      \begin{enumerate}
\item $\Delta \ell=1$ scattering processes: \\
	standard  $s$-channel processes :\\ $ \ell~ N_1  \leftrightarrow  d~\bar{u}$, \quad $\bar{\ell} ~ N_1  \leftrightarrow  \bar{d}~ u$,    \quad 
    
   standard $t$-channel processes :\\ $N_1 ~u  \leftrightarrow d~\bar{\ell} $, \quad $N_1 ~\bar{u}  \leftrightarrow \bar{d}~ \ell$, \quad 

 $N_1~ d  \leftrightarrow u ~\ell$, \quad $N_1 ~\bar{d}  \leftrightarrow \bar{u} ~\bar{\ell} $  \quad 

standard processes involving gauge boson ($V_{\mu}= W^i_{\mu}$ and $B_{\mu}$ ) \\ 

$N_1 ~H \xleftrightarrow{s} V_{\mu} ~\ell$, \quad $N_1 H^{\dagger}  \xleftrightarrow{s} \bar V_{\mu} ~\bar{\ell}$, \quad \\

$N_1 ~\ell  \xleftrightarrow{s} V_{\mu} ~H$, \quad $N_1 \bar{\ell} \xleftrightarrow{s} \bar V_{\mu} ~H^{\dagger}$, \quad 

$N_1~V_{\mu}  \xleftrightarrow{t} H~ \ell$, \quad $N_1~\bar V_{\mu}  \xleftrightarrow{t} \bar H~ \bar \ell$ \quad 

$N_1~V_{\mu}  \xleftrightarrow{u} H~ \ell$, \quad $N_1~\bar V_{\mu}  \xleftrightarrow{u} \bar H~ \bar \ell$ \quad 

$N_1 ~ \bar H  \xleftrightarrow{t} V_{\mu} ~\ell$, \quad $N_1 {H}  \xleftrightarrow{t} \bar V_{\mu} ~\bar{\ell}$, \quad 

$N_1 ~\ell  \xleftrightarrow{t} V_{\mu} ~\bar H$, \quad $N_1 \bar{\ell}  \xleftrightarrow{t} \bar V_{\mu} ~{H}$, \quad 

Non-standard scattering processes: \\
    $\ell ~H  \xleftrightarrow{s} \phi ~\psi$, \quad  $  \phi ~\psi   \xleftrightarrow{s} \ell ~H  $,  \quad   
    
    $\ell ~H  \xleftrightarrow{s} \phi^{\dagger}~ \bar{\psi}$, \quad 
    $  \phi^{\dagger}~ \bar{\psi}  \xleftrightarrow{s}   \ell ~H        $  \quad   \\
   $\ell ~\phi  \xleftrightarrow{t} H ~\psi$,\quad $ H ~\psi   \xleftrightarrow{t} \ell ~\phi  $, \quad \\
   $\bar \ell ~\psi  \xleftrightarrow{t} \bar H ~{\phi}$, \quad  $  \bar H ~{\phi}   \xleftrightarrow{t}   \bar \ell ~\psi$ \quad \\
   $\bar{\ell}~ \phi  \xleftrightarrow{t} {H^{\dagger} } ~\psi$, \quad  $    {H^{\dagger} } ~\psi   \xleftrightarrow{t}\bar{\ell}~ \phi  $,  \quad 
    \\
     $  H ~\phi  \xleftrightarrow{t} \ell ~\psi  $, \quad  $\bar \ell ~ \bar \psi  \xleftrightarrow{t} \bar H ~\bar \phi$, \quad \\
    $\phi~ \psi  \xleftrightarrow{s+t} \phi ~\bar{\psi}$,  \quad
     $\phi~ \bar{\psi}  \xleftrightarrow{s+t} \phi ~\psi$,  \quad
     $\phi~ \phi  \xleftrightarrow{t} \psi ~ \bar{\psi}$,   \quad 
    $\phi ~\phi  \xleftrightarrow{t} \psi ~ \psi$, \quad  $\phi ~\phi  \xleftrightarrow{t} \bar{\psi} ~\bar{\psi}$; 
\item $\Delta \ell=2$ scattering processes:  $\ell ~\ell  \xleftrightarrow{t+u} H^{\dagger} ~H^{\dagger}$, \quad \quad $H ~\ell  \xleftrightarrow{s+t} H^{\dagger} ~\bar{\ell}$, \quad 
\\
    and all the reverse reactions.
\end{enumerate}

As we can identify that some of the decay and scattering processes mention above are new, while others are studied in standard thermal leptogenesis,~\cite{Xing:2011zza},\cite{Buchmuller:2004nz},\cite{Giudice:2003jh}.

\section{\texorpdfstring{Decay Widths of $N_1$}{Decay Widths of N1}}
The lightest right-handed neutrino $N_1$ can decay into two modes: $N_1 \to lH$ and $N_1 \to \psi \phi$. The decay widths of two modes $N_1 \to lH$ and $N_1 \to \psi \phi$ are given as follows
 \begin{equation}
    \Gamma_{N_1 \to \ell  H } = \frac{(Y^{N^\dagger}Y^N)_{11}}{8 \pi ~m_{N_1}^3} \left( m^2_{N_1}  - m_{H}^2 \right)^2,~{\rm and }
    \label{eq:dN1tolh}
\end{equation}
\begin{equation}
     \Gamma_{N_1 \to \phi ~\psi} = \frac{|y_1|^2}{16 \pi ~m_{N_1}^3} \left[ m^2_{N_1} + m_\psi^2 - m_\phi^2 \right] \sqrt{ (m_{N_1}^2 - m_\phi^2 - m_\psi^2)^2 - 4 ~m_\phi^2 ~m_\psi^2 }
\end{equation}
where $y_1$ and $Y^{N}$ denote the Yukawa couplings, and 
$m_{\phi}$, $m_{\psi}$, $m_{N_1}$, and $m_{H}$ represent the masses of the dark matter particles $\phi$ and $\psi$, the lightest right-handed neutrino $N_1$, and the Higgs boson, respectively.

\section{Scattering cross sections}
Several scattering processes contribute to the lepton asymmetry. Here, we provide the cross-section of one of the $s$ channel $2\to 2$ scattering process $\phi\,\psi \xrightarrow{s} \ell\,H $ as: 
\begin{align}
\sigma_{\phi \psi \to \ell H}
= \frac{|y_1|^2\, K_{11}\,(s-m_H^2)^2}
{64\pi\,\sqrt{s(s-4m_\psi^2)}\,
\left[(s-m_{N_1}^2)^2 + m_{N_1}^2\,\Gamma_1^2\right]} \ , 
\end{align}
where, \(\sqrt{s}\) denotes the center-of-mass energy of the process, \(m_\psi\) is DM mass, \(m_H\) is the Higgs boson mass,  \(m_{N_1}\) is the mass of the lightest right-handed neutrino \(N_1\), and \(\Gamma_1\) is its total decay width. In addition, the quantity \(K_{11} = (Y^{N}Y^{N\dagger})_{11}\) arises from the see-saw Yukawa matrix and \(y_1\) is the Yukawa coupling associated with the \(\phi\)–\(\psi\)–\(N_1\) interaction.



\section{\texorpdfstring{Thermal average of decay width}{Thermal average of decay width}}
Consider a particle $X$ that decays into particles $a$ and $b$. 
Let $\Gamma_X$ denote the decay width of the particle $X$. 
The thermally averaged decay width of $X$ is given by 
\begin{align}
\gamma_X &=  \int d\Pi_{X} e^{-E_X/T}  \int d\Pi_{a} d\Pi_{b}  (2 \pi )^4 \delta^{4}(p_X-p_a-p_b)|M_{X \to a b}|^2 \nonumber \\
    &= n_X^{\mathrm{eq}} \,
      \frac{K_1(z)}{K_2(z)} \,
      \Gamma_{X \to a b},
    \label{eq:thdX}
\end{align}
where $n_X^{\mathrm{eq}}$ is the equilibrium number density of $X$, and $\Gamma_{X \to a b}$ is the decay width of $X \to a b$  and $K_1(z)$ and $K_2(z)$ are the modified Bessel functions of the second kind of order one and two, respectively. The modified Bessel functions of the second kind of order one $K_1(z)$ and two $K_2(z)$ are as follows:
\begin{align}
    K_1(z) &= \frac{1}{z}
    \int_{z}^{\infty} e^{-y}\sqrt{y^2 - z^2}\, \mathrm{d}y , ~{\rm and} \\
    K_2(z) &= \frac{1}{z^2}
    \int_{z}^{\infty} y\, e^{-y}\sqrt{y^2 - z^2}\, \mathrm{d}y,
\end{align}
respectively.



\section{\texorpdfstring{Thermal average cross-section of $2 \to 2$ scattering processes}{Thermal average of 2 -> 2 scattering processes}}
\label{sec:apendez:one}
 
The thermal average cross-section \(\gamma\) for a \(2 \to 2\) scattering process with \(a\) and \(b\) being incoming particles, is given by
\begin{align}
    \gamma^{ab} = \frac{T}{64 \pi^4} \int_{s_{\text{min}}}^\infty \mathrm{d}s\, \sqrt{s} \, \hat{\sigma}^{ab}(s)\, K_1\left( \frac{\sqrt{s}}{T} \right),
    \label{eq:apendex:thermal:average}
\end{align}
where:
\begin{itemize}
    \item \(T\) is the temperature,
    \item \(\sqrt{s}\) is the center-of-mass energy,
    \item \(K_1\) is the modified Bessel function of the second kind of order one,
    \item \(\hat{\sigma}(s)\) is the \emph{reduced cross-section}, defined as
    \begin{align}
        \hat{\sigma}^{ab}(s) = 2s\, \sigma^{ab}(s)\, \lambda\left(1, \frac{m_{a}^2}{s}, \frac{m_{b}^2}{s}\right),
    \end{align}
    where \(\sigma^{ab}(s)\) is the cross-section for the process, and \(\lambda(x, y, z) = x^2 + y^2 + z^2 - 2(xy + yz + zx)\).
    \item \(m_{a}, m_{b}\) are the masses of the incoming particles.
\end{itemize}
The lower limit of integration, \(s_{\text{min}} = \max[(m_{a_1} + m_{a_2})^2, (m_{a_3} + m_{a_4})^2]\), ensures that the process is kinematically allowed.

\textbf{Note:} For scattering processes that violate $C\!P$ asymmetry, the definition of thermally averaged cross-section of scattering process  in Eq.~\ref{eq:apendex:thermal:average} slightly modifies to Eq.~\ref{eq:apendex:thermal:averageep}. In this case, the reduced cross section $\hat{\sigma}$ is replaced with $\epsilon\hat {\sigma}$ (or $\hat{\sigma} \to \epsilon\hat {\sigma} $), where $\epsilon$ denotes the $C\!P$-asymmetry parameter. Finally, the thermally averaged cross-section of scattering processes that violate $C\!P$ asymmetry $\Theta^{ab}_{\epsilon_{\alpha}}$ is given by 
\begin{align}
    \Theta^{ab}_{\epsilon_{\alpha}} = \frac{T}{64 \pi^4} \int_{s_{\text{min}}}^\infty \mathrm{d}s\, \sqrt{s} \, ~\epsilon_a\hat{\sigma}^{ab}(s)\,~ K_1\left( \frac{\sqrt{s}}{T} \right),
    \label{eq:apendex:thermal:averageep}
\end{align}
where other details remain same as in Eq. ~\ref{eq:apendex:thermal:average}.


\bibliographystyle{apsrev4-2}
\bibliography{ref1}
\end{document}